\documentclass[12pt]{article}

\usepackage[toc,page]{appendix}
\usepackage{slashed}
\usepackage[sumlimits,intlimits,namelimits]{amsmath}
\usepackage[english]{babel}
\usepackage{graphicx}
\usepackage{hyperref}
\usepackage{cite}

\newcommand{\ba}{\begin{eqnarray}}
\newcommand{\ea}{\end{eqnarray}}

\newcommand{\ice}[1]{\relax}

\usepackage{color}
\usepackage{amsmath}
\usepackage{booktabs}
\usepackage{subfigure}

\DeclareMathOperator{\Tr}{Tr}

\def\Li{\mbox{Li}}

\begin{document}

\begin{titlepage}

\begin{flushright}
PSI-PR-24-07
\end{flushright}
\vspace{1.2cm}
\begin{center}
	  	  { \Large\bf QCD corrections to the Darwin coefficient in
	  	  inclusive semileptonic $B\rightarrow X_u \ell \bar{\nu}_\ell$ decays}
\end{center}
\vspace{0.5cm}
\begin{center}
{\sc Daniel Moreno} \\[0.2cm]
{\sf \it Paul Scherrer Institut, CH-5232 Villigen PSI, Switzerland}
\end{center}

\vspace{0.8cm}
\begin{abstract}\noindent
 In this paper we compute $\alpha_s$ corrections to the matching coefficients of the dimension six operators
 in the Heavy Quark Expansion of the inclusive semileptonic heavy hadron decay rate and leptonic invariant mass spectrum with a massless quark and both a massive or massless lepton in the final state, analytically.
 The obtained results can be applied to the inclusive semileptonic $B\rightarrow X_u \ell \bar{\nu}_\ell$ ($\ell = e,\,\mu,\,\tau$) and
 $D\rightarrow X \ell \bar{\nu}_\ell$ ($\ell = e,\,\mu$) decays.
 The main application of our results is the background subtraction of the $B\rightarrow X_u \ell \bar{\nu}_\ell$ decay
 in the measurement of the $B\rightarrow X_c\ell \bar{\nu}_\ell$ decay, which is important for the precise extraction of $V_{cb}$ and $R(D^{(*)})$. They also play a role in the computation of lifetimes of heavy hadrons.
\end{abstract}

\end{titlepage}

\section{Introduction} 
\label{sec:Intro}
The heavy quark expansion (HQE)~\cite{Shifman:1987rj,Eichten:1989zv,Isgur:1989vq,Grinstein:1990mj}, as the standard tool
for the description of inclusive heavy hadron decay rates and distributions~\cite{Chay:1990da,Bigi:1992su,Bigi:1993fe,Blok:1993va,Manohar:1993qn}, has been subject to intensive progress since its birth around thirty years ago and it is expected to improve
further in the near future motivated by the precise determination of
$|V_{cb}|$~\cite{Bordone:2021oof,Bernlochner:2022ucr,Finauri:2023kte,Fael:2022wfc,Bigi:2023cbv}, obtaining precise predictions for
$B$-hadron lifetimes~\cite{HFLAV:2022esi,Lenz:2014jha,Cheng:2018rkz,Lenz:2022rbq,Gratrex:2023pfn,Piscopo:2023jnu,Mannel:2023zei,Albrecht:2024oyn}, exploring its applicability and non-perturbative effects
in $D$-decays~\cite{Neubert:1996we,Mannel:2021uoz,King:2021xqp,King:2021jsq,Gratrex:2022xpm,Cheng:2023jpz,Dulibic:2023jeu} and
exploring semitauonic decays from an inclusive perspective~\cite{Belle-II:2023aih} in light of the $R(D^{(*)})$ anomaly present in the exclusive
channels~\cite{LHCb:2017rln,Belle:2019rba,Bernlochner:2021vlv,Ligeti:2014kia,Ligeti:2021six,Rahimi:2022vlv}.

The main assumption in the construction of the HQE is that the heavy hadron momentum $p_{H_Q} = p_Q + k$ is equal to the heavy quark momentum $p_Q$ up to fluctuations $k$ of the order of the QCD hadronization scale $\Lambda_{\rm QCD}$. In other words, due to its large mass $m_Q$ the heavy quark inside the heavy hadron is almost on shell.

By taking advantage of the fact that $m_Q\gg\Lambda_{\rm QCD}$ one can construct an operator product expansion, the so-called HQE.
As a result one obtains the decay width and distributions as a power expansion in $\Lambda_{\rm QCD}/m_Q$, whose coefficients have
a perturbative expansion in the strong coupling $\alpha_s(m_Q)$.
A systematic improvement is possible by calculating higher orders in the two expansion parameters. The HQE has
proven to be a reliable tool to describe inclusive $B$ ($m_Q=m_b$) and, to some extent, $D$ ($m_Q=m_c$) decays.

In this paper we consider higher order corrections in the HQE of the inclusive semileptonic $B \rightarrow X_u \ell \bar{\nu}_\ell$ decay rate and
distribution in the dilepton invariant mass $q^2$, with $\ell=e,\,\mu,\,\tau$.
A precise experimental measurement of the differential rate of $B\rightarrow X_u \ell \bar\nu_\ell$
relies on experimental cuts to suppress the overwhelming contamination of the $B\rightarrow X_c \ell \bar\nu_\ell$ decay.
From the theory side, these cuts have the troublesome consequence that on the remaining phase space, where the decay is measured,
perturbation theory breaks down and it is not possible to use the HQE. A theoretical description in such a region relies on non-perturbative methods involving shape functions~\cite{Neubert:1993ch}, which is important to extract $|V_{ub}|$ from
inclusive decays. Contrarily, a precise theoretical description of the $B\rightarrow X_u \ell \bar\nu_\ell$ decay in the region where
the HQE is applicable is important for reliably modeling this channel as a background in the measurement of
the $B\rightarrow X_c \ell \bar\nu_\ell$ decay, used for the precise extraction
of $|V_{cb}|$ from inclusive decays~\cite{Belle:2021idw,Bordone:2021oof,Bernlochner:2022ucr,Finauri:2023kte} ($\ell=e,\,\mu$) and
$R(D^{(*)})$ ($\ell=\tau$)~\cite{Ligeti:2021six}.

The total rate of $B\rightarrow X_u \ell \bar\nu_\ell$ is the most inclusive quantity and
it can be computed within the HQE with impact on the predictions for $B$-hadron lifetimes, even though this decay channel
is CKM (Cabibbo–Kobayashi–Maskawa) suppressed, and therefore the impact is low. However, its precise
determination is very appealing in light of the recent preliminary measurement by the Belle II collaboration of the ratio
$\Gamma(B\rightarrow X_u \ell \bar{\nu}_\ell)/\Gamma(B\rightarrow X_c \ell \bar{\nu}_\ell)$ \cite{cao:talk} which allows to extract the ratio $|V_{ub}/V_{cb}|$~\cite{Alberti:2014yda}.
This ratio also enters as a normalization factor in the branching ratio
of $B \rightarrow X_s \gamma$ and $B\rightarrow X_s \ell^{+}\ell^{-}$~\cite{Huber:2020vup}.

Also the inclusive semitauonic decay $B\rightarrow X_u \tau \bar \nu_\tau$ could be measured in the near future by Belle II~\cite{Ligeti:2014kia,Ligeti:2021six}. 
For example, the collaboration has already set the first bound on a $b\rightarrow u \tau \bar{\nu}_\tau$ mediated
decay~\cite{Belle:2015qal}.

The expressions obtained for the $B\rightarrow X_u \ell \bar\nu_\ell$ ($\ell=e,\,\mu$) decay can be directly applied to the inclusive
semileptonic $D\rightarrow X \ell \bar{\nu}_\ell$ decay. Unlike in the case of bottom, for charm the results can be applied to the
CKM favoured decay channel $c\rightarrow s \ell \bar{\nu}_\ell$ and therefore they will have a larger impact.
The $q^2$-spectrum in $D$ mesons also offers an opportunity to test the HQE for $D$ decays beyond the
total rate.

The current status of the HQE for inclusive semileptonic decay rates and distributions (we mainly refer to the $q^2$-distribution)
is the following:

\begin{itemize}
 \item $B \rightarrow X_c \ell \bar{\nu}_\ell$:
 The leading power coefficient is known for the total rate and a variety of moments
 at next-to-next-to-next-to-leading order (N$^3$LO)~\cite{Nir:1989rm,vanRitbergen:1999gs,Pak:2008qt,Pak:2008cp,Fael:2020tow,Czakon:2021ybq,Fael:2022frj,Egner:2023kxw}
 and at next-to-next-to-leading order (N$^2$LO)~\cite{Ho-kim:1983klw,Czarnecki:1994bn,Jezabek:1996db,Jezabek:1997rk,Biswas:2009rb} in the case of a
 massless and massive lepton in the final state, respectively.
The $1/m_b^2$ and $1/m_b^3$ corrections are known for the width and some distributions
at next-to-leading order (NLO)~\cite{Balk:1993sz,Koyrakh:1993pq,Falk:1994gw,Gremm:1996df,Alberti:2013kxa,Ligeti:2014kia,Mannel:2014xza,Mannel:2015jka,Mannel:2017jfk,Mannel:2019qel,Colangelo:2020vhu,Mannel:2021zzr,Rahimi:2022vlv,Moreno:2022goo} in both cases. Finally, the $1/m_b^4$ and $1/m_b^5$ corrections are
known at leading order (LO)~\cite{Dassinger:2006md,Mannel:2010wj,Mannel:2023yqf} in the massless lepton case.
 \item $B \rightarrow X_u \ell \bar{\nu}_\ell$:
 In the case of a massless quark in the final state (with either massive or massless leptons) the coefficients up to order $1/m_b^2$ can be straightforwardly obtained from the $B \rightarrow X_c \ell \bar{\nu}_\ell$ decay by taking the limit $m_c\rightarrow 0$\footnote{This statement is generally true for the decay rate and the $q^2$ spectrum, but not for other distributions like the lepton energy spectrum~\cite{Ligeti:2021six}. Nevertheless, methods have been developed and applied to take the limit $m_c\rightarrow 0$ also for
 such distributions up to order $\alpha_s/m_b^2$~\cite{Capdevila:2021vkf}.}.
 This statement requires some caveats concerning the N$^3$LO corrections as they have been
 estimated by taking an expansion at small $\delta = 1 - m_c/m_b$~\cite{Fael:2020tow,Fael:2022frj,Czakon:2021ybq} with $\delta \rightarrow 1$. However the expansion shows a good convergence even for $\delta \rightarrow 1$~\cite{Fael:2023tcv}.
 The N$^3$LO corrections have been also computed in the leading-color approximation~\cite{Chen:2023dsi}, the later including
 results for the $q^2$-spectrum. The subset of five-loop diagrams containing closed fermionic
 loops have been computed without any approximations in~\cite{Fael:2023tcv}.
 Starting at order $1/m_b^3$ onward it is not possible to extrapolate $m_c\rightarrow 0$, which is related to the
 appearance of four-quark operators in the operator basis of the HQE. At $1/m_b^3$ the coefficients of the total rate for
 the two-quark operators and four-quark operators are known at LO~\cite{Lenz:2020oce,Mannel:2020fts,MorenoTorres:2020xir} and NLO~\cite{Beneke:2002rj,Lenz:2013aua}, respectively.
\end{itemize}

This paper is a follow-up of Refs.~\cite{Mannel:2021zzr,Moreno:2022goo}, where the coefficients of the $1/m_b^3$ terms were
computed at NLO for the $B\rightarrow X_c \ell \bar{\nu}_\ell$ decay rate and $q^2$-distribution,
i.~e. for the case of a massive quark in the final state. This includes the Darwin and
spin-orbit operator coefficients, where the latter is related to coefficients of lower orders in the $\Lambda_{\rm QCD}/m_b$ expansion
by reparametrization invariance~\cite{Manohar:2010sf,Becher:2007tk}.
We extend our previous calculation to the $B\rightarrow X_u \ell \bar{\nu}_\ell$ decay, where the final-state quark is massless. We consider both cases, a massless and a massive lepton in the final state. 
In this case, the operator basis of the HQE at $1/m_b^3$ also includes four-quark operators.
As already mentioned, for the Darwin coefficient the limit $m_c\rightarrow 0$ can not be straightforwardly obtained from the
results of these papers since for $m_c=0$ the coefficient is infrared singular, pointing out the operator mixing of the Darwin operator with four-quark operators under renormalization. 
In turn, we also obtain results for the coefficients of the four-quark operators of the differential rate at NLO which, to
the best of my knowledge, have never been presented in previous studies.
Moments of the spectrum with arbitrary cuts can be obtained by integrating the differential rate with the corresponding
weight function in the desired range.

We provide a Mathematica file \textit{``coefbutv.nb''} containing analytical results for the NLO coefficients of the Darwin operator and four-quark operators appearing at $1/m_Q^3$ for both, 
the total width and the $q^2$-spectrum of the $B\rightarrow X_u \ell \bar{\nu}_\ell$ decay ($\ell=e,\,\mu,\,\tau$).

We organize the paper as follows. In Sec.~\ref{sec:HQEIDHH} we give our main definitions for
the HQE of inclusive heavy hadron decays. In
Sec.~\ref{sec:HQE} we outline the calculation with Secs.~\ref{sec:4qO} and \ref{sec:darwin} devoted to the computation of the four-quark operator coefficients and the Darwin operator coefficient, respectively.
In Sec.~\ref{sec:ev} we discuss the use of evanescent operators.
Finally, we discuss the impact of our results in Sec.~\ref{sec:disc}.

\section{HQE for inclusive decays of heavy flavoured hadrons}
\label{sec:HQEIDHH}

This section provides a brief overview of the theoretical framework employed for the computation of inclusive semileptonic decays of heavy hadrons and outlines key definitions.

When the momentum transfer is considerably lower than the W-boson mass, the heavy quark decay
$Q\rightarrow q \ell \bar\nu_\ell$, which is mediated by a charged current interaction, is described by an effective Fermi Lagrangian
\begin{equation}
  \label{eq:FermiLagr}
{\cal L}_{\rm eff} = 2\sqrt{2}G_F  V_{qQ}(\bar{Q}_L \gamma_\mu q_L)
(\bar{\nu}_{\ell, L} \gamma^\mu \ell_L) + {\rm h.c.} \, ,
\end{equation}
where the subscript $L$ stands for left-handed fermion fields, $G_F$ is the Fermi constant
and $V_{qQ}$ is the corresponding element of the CKM matrix. We denote the heavy quark mass by $m_Q$
and define the dimensionless quantity $\eta = m_\ell^2/m_Q^2$, where $m_\ell$ is the lepton mass. The light quark $q$ in the final state is considered to be massless.

The inclusive decay rate of the heavy hadron $H_Q$ made of the heavy quark $Q$
is then obtained, by virtue of the optical theorem, from
the imaginary part of the forward hadronic matrix element of the transition operator
${\cal T}$
\begin{equation}\label{eq:trans_operator}
{\cal T} = i \int d^D x\,
T\left\{ {\cal L}_{\rm eff} (x) {\cal L}_{\rm eff} (0) \right\} \, ,
\quad \Gamma (H_Q \to X_q \ell \bar{\nu}_\ell)
= \frac{1}{M_{H_Q}} \text{Im }\langle H_Q|{\cal T} |H_Q \rangle \,,
\end{equation} 
where the heavy hadron is represented by a full QCD state $|H_Q \rangle$ with mass $M_{H_Q}$, velocity $v$ and momentum $p_{H_Q} = M_{H_Q} v$. We regularize both ultraviolet and infrared divergences in Eq.~(\ref{eq:trans_operator}) in standard dimensional
regularization with $D= 4-2\epsilon$ spacetime dimensions.

Since $m_Q\gg \Lambda_{\rm QCD}$ the above equation contains contributions that can be computed within
perturbation theory. These contributions can be factorized from the non-perturbative ones by using the
HQE, where the imaginary part of ${\cal T}$ is matched to an expansion in
$\Lambda_{\rm QCD}/m_Q$ by employing local operators in heavy quark effective theory (HQET)~\cite{Mannel:1991mc,Manohar:1997qy}
\begin{eqnarray}
	\label{eq:HQE-1}
	\mbox{Im}\, \mathcal{T} &=& \Gamma^0 |V_{qQ}|^2
	\bigg( C_0 \mathcal{O}_0
	+ C_v \frac{\mathcal{O}_v}{m_Q}
	+ C_\pi \frac{\mathcal{O}_\pi}{2m_Q^2}
	+ C_G \frac{\mathcal{O}_G}{2m_Q^2}
	+ C_D \frac{\mathcal{O}_D}{4m_Q^3}
	+ C_{LS} \frac{\mathcal{O}_{LS}}{4m_Q^3}
	\nonumber
	\\
	&&
	+ C_1^{hl} \frac{\mathcal{O}_1^{hl}}{4m_Q^3}
	+ C_2^{hl} \frac{\mathcal{O}_2^{hl}}{4m_Q^3}
	+ C_3^{hl} \frac{\mathcal{O}_3^{hl}}{4m_Q^3}
	+ C_4^{hl} \frac{\mathcal{O}_4^{hl}}{4m_Q^3}
    + C_5^{hl} \frac{\mathcal{O}_5^{hl}}{4m_Q^3}
    + C_6^{hl} \frac{\mathcal{O}_6^{hl}}{4m_Q^3}
	\bigg)\,,
\end{eqnarray}
where $\Gamma^0 = G_F^2 m_Q^5/(192 \pi^3)$ and $C_i=C_i(\eta)$ are the matching coefficients, which can be computed as a perturbative expansion in the strong coupling $\alpha_s (\mu)$. In the text we will refer to the different orders in the $\Lambda_{\rm QCD}/m_Q$ expansion as leading power, next-to-leading power and so on. Similarly we will refer to the different orders in the $\alpha_s(\mu)$ expansion as LO, NLO and so on. Finally $\mathcal{O}_i$ denotes the HQET operators, which we list below

\begin{alignat}{2}
\mathcal{O}_0 &= \bar h_v h_v   &&\mbox{(leading power operator)}\,, \\
\mathcal{O}_v &= \bar h_v v\cdot \pi h_v &&\mbox{(EOM operator)}\,, \\
\mathcal{O}_\pi &= \bar h_v \pi_\perp^2 h_v &&\mbox{(kinetic operator)}\,, \label{mupi} \\
\mathcal{O}_G &= \frac{1}{2}\bar h_v [\gamma^\mu, \gamma^\nu] \pi_{\perp\,\mu}\pi_{\perp\,\nu}  h_v &&\mbox{(chromomagnetic operator)}\,, \label{muG} \\
\mathcal{O}_D &= \bar h_v[\pi_{\perp\,\mu},[\pi_{\perp}^\mu , v\cdot \pi]] h_v && \mbox{(Darwin operator)}\,,  \label{rhoD}\\
\mathcal{O}_{LS} &= \frac{1}{2}\bar h_v[\gamma^\mu,\gamma^\nu]\{ \pi_{\perp\,\mu},[\pi_{\perp\,\nu}, v\cdot \pi] \} h_v && \mbox{(spin-orbit operator)}\,, \label{rhoLS} \\
\mathcal{O}_1^{hl} &= (\bar h_v \gamma_{\mu} P_L q)(\bar q \gamma^\mu P_L h_v) && \mbox{(vector singlet operator)}\,, \label{o1hl} \\
 \mathcal{O}_2^{hl} &= (\bar h_v P_L q)(\bar q P_R h_v) && \mbox{(scalar singlet operator)}\,,\label{o2hl} \\
\mathcal{O}_3^{hl} &= (\bar h_v \gamma_{\mu} P_L T^a q)(\bar q \gamma^\mu P_L T^a h_v) && \mbox{(vector octet operator)}\,,
 \label{o3hl} \\
\mathcal{O}_4^{hl} &= (\bar h_v P_L T^a q)(\bar q P_R T^a h_v) && \mbox{(scalar octet operator)}\,,
 \label{o4hl} \\
\mathcal{O}_5^{hl} &= (\bar h_v \gamma_\mu \gamma_\nu P_L T^a q)(\bar q \gamma^\mu \gamma^\nu P_R T^a h_v) && \mbox{(rank-2 tensor octet operator)}\,,
 \label{o5hl} \\
 \mathcal{O}_6^{hl} &= (\bar h_v \gamma_\mu \gamma_\nu \gamma_\alpha P_L T^a q)(\bar q \gamma^\mu \gamma^\nu \gamma^\alpha P_L T^a h_v) \quad && \mbox{(rank-3 tensor octet operator)}\,,
 \label{o6hl}
\end{alignat}
with $P_{R/L} = (1\pm \gamma_5)/2$ being the right/left handed projectors,
$\pi_\mu = i D_\mu = i\partial_\mu +g_s A_\mu^a T^a$ the QCD covariant derivative,
$a^\mu_\perp = a^\mu - v^\mu (v\cdot a)$, and $h_v$ the HQET field whose momentum is of the order of $\Lambda_{\rm QCD}$ and whose dynamics is determined by the HQET Lagrangian~\cite{Manohar:1997qy}.

Since the quark $q$ is considered to be massless it remains a dynamical degree of freedom in the effective theory and therefore
it must be used in the construction of the operator basis of the HQE. This degree of freedom shows up first as
the four-quark operators in Eqs.~(\ref{o1hl})-(\ref{o6hl}). We use the index ``hl'' to denote such operators which we will also refer as
heavy-light operators.

Note that in $D=4$ the operators $\mathcal{O}_{5,6}^{hl}$ can be reduced to the operators $\mathcal{O}_{3,4}^{hl}$.
This is no longer true if we work in dimensional regularization where the operator basis is formally infinite dimensional. In such a case four-quark operators with arbitrarily long strings of gamma matrices contracted in two different fermion lines must be included as
elements of the basis. Nevertheless, at a fixed order in perturbation theory only a finite number of four-quark operators appear.
To the order we are working on, only the ones explicitly written above are relevant.
Despite of this, it is also possible to make a connection of a $D$-dimensional operator basis like the one we have chosen to
the four-dimensional operator basis. However, this requires the introduction of evanescent
operators. We will devote Sec.~(\ref{sec:ev}) to a detailed discussion on the topic.

Finally, operators which are of higher dimension after using the EOM of HQET have been neglected in Eq.~(\ref{eq:HQE-1}).
However, the matching calculation is done off shell, and such operators affect the extraction of the Darwin operator coefficient, i.e. the projector to the corresponding coefficient. Only after the matching calculation, the operators are removed by
using the EOM.

Since we are also interested in the spectrum in the dilepton invariant mass $q^2$ we perform the matching at the differential level~\cite{Mannel:2021ubk}. This can be achieved on the one hand by writing the HQE of the $\mbox{Im }\mathcal{T}$ in differential form
\begin{eqnarray}
	\label{eq:HQE-1-dq}
	\mbox{Im}\, \mathcal{T} &=& \Gamma^0 |V_{qQ}|^2 \int_{\eta}^{1} dr
	\bigg( \mathcal{C}_0 \mathcal{O}_0
	+ \mathcal{C}_v \frac{\mathcal{O}_v}{m_Q}
	+ \mathcal{C}_\pi \frac{\mathcal{O}_\pi}{2m_Q^2}
	+ \mathcal{C}_G \frac{\mathcal{O}_G}{2m_Q^2}
	+ \mathcal{C}_D \frac{\mathcal{O}_D}{4m_Q^3}
	+ \mathcal{C}_{LS} \frac{\mathcal{O}_{LS}}{4m_Q^3}
	\nonumber
	\\
	&&
	+ \mathcal{C}_1^{hl} \frac{\mathcal{O}_1^{hl}}{4m_Q^3}
	+ \mathcal{C}_2^{hl} \frac{\mathcal{O}_2^{hl}}{4m_Q^3}
	+ \mathcal{C}_3^{hl} \frac{\mathcal{O}_3^{hl}}{4m_Q^3}
	+ \mathcal{C}_4^{hl} \frac{\mathcal{O}_4^{hl}}{4m_Q^3}
    + \mathcal{C}_5^{hl} \frac{\mathcal{O}_5^{hl}}{4m_Q^3}
    + \mathcal{C}_6^{hl} \frac{\mathcal{O}_6^{hl}}{4m_Q^3}
	\bigg)\,,
\end{eqnarray}
by defining the coefficients of the differential rate $\mathcal{C}_i(r,\eta)$ through the coefficients of the total rate
\begin{eqnarray}
 C_i(\eta) &=& \int_{\eta}^{1} dr\, \mathcal{C}_i(r,\eta)\,,
 \label{Citot}
\end{eqnarray}
where $r = q^2/m_Q^2$ ($\eta \le r \le 1$) is the dilepton invariant mass normalized to the heavy quark mass. And on the other hand, by using a dispersion representation defined in dimensional regularization~\cite{Groote:1999zp}
for the lepton-antineutrino loop on the QCD side, i.e in Eq.~(\ref{eq:trans_operator}). Note that the use of such a representation is
always possible because the leptonic part is not affected by QCD corrections and therefore appears factorized from the hadronic part.
For a massive lepton and a massless antineutrino the dispersive relation reads
\begin{eqnarray}
\label{spectrum1}
 &&i\int\frac{d^D k}{(2\pi)^D}
 \frac{-\Tr(\gamma^\sigma P_L i(\slashed k + \slashed \ell + m_\ell)\gamma^\rho P_L i\slashed k)}{k^2((k+\ell)^2 - m_\ell^2)}
 \nonumber
 \\
  &&\quad\quad =
 \int_{m_\ell^2}^{\infty} d(q^2) \frac{1}{q^2 - \ell^2-i\eta}
 \frac{1}{(4\pi)^{D/2}}\frac{\Gamma(D/2 - 1)}{\Gamma(D - 2)} \frac{D-2}{D-1}
 \nonumber
 \\
 &&
 \quad\quad\quad \times
 (q^2)^{D/2 - 2} \bigg(1- \frac{m_\ell^2}{q^2}\bigg)^{D - 2}
   \bigg[
   \bigg( 1 + \frac{D}{D-2}\frac{ m_\ell^2 }{q^2} \bigg) \ell^\rho \ell^\sigma
  - \bigg(  1 + \frac{1}{D-2}\frac{m_\ell^2}{q^2} \bigg) \ell^2 g^{\rho\sigma}
  \bigg]\,,
\end{eqnarray}
where $\ell$ is the four-momentum flowing through the leptons.
Note that there is no need for subtractions in the dispersion relation above since singularities are regularized by dimensional regularization. 
Also observe that by employing the dispersion representation above the leptonic part becomes, at the differential level,
a massive propagator of mass $q$ and the dependence on $m_\ell$ factorizes from the hadronic part. In particular
it implies that the differential rate is a polynomial in $\eta$ of degree three.
Therefore, the master integrals needed for the computation of the differential rate are the same in both cases,
for a massive or a massless lepton
in the final state. The corresponding master integrals can be found in \cite{Mannel:2021ubk}.

Exchanging the leading term operator $\mathcal{O}_0$ in Eq.~(\ref{eq:HQE-1-dq}) by the local QCD operator
$\bar Q \slashed v Q$ is advantageous, since its forward hadronic matrix element (see Eq.~\eqref{eq:hadrforwardv}) is completely normalized. To that end,
we need the HQE of the $\bar Q \slashed v Q$ operator up to the desired order
\begin{eqnarray}
 \bar Q \slashed v Q &=&
 \mathcal{O}_0
 + \tilde{C}_v \frac{\mathcal{O}_v}{m_Q}
 + \tilde C_\pi \frac{\mathcal{O}_\pi}{2m_Q^2}
 + \tilde C_G \frac{\mathcal{O}_G}{2m_Q^2}
 + \tilde C_D \frac{\mathcal{O}_D}{4m_Q^3}
 + \tilde C_{LS} \frac{\mathcal{O}_{LS} }{4m_Q^3}
 \nonumber
 \\
 &&
	+ \tilde C_1^{hl} \frac{\mathcal{O}_1^{hl}}{4m_Q^3}
	+ \tilde C_2^{hl} \frac{\mathcal{O}_2^{hl}}{4m_Q^3}
	+ \tilde C_3^{hl} \frac{\mathcal{O}_3^{hl}}{4m_Q^3}
	+ \tilde C_4^{hl} \frac{\mathcal{O}_4^{hl}}{4m_Q^3}
	+ \tilde C_5^{hl} \frac{\mathcal{O}_5^{hl}}{4m_Q^3}
	+ \tilde C_6^{hl} \frac{\mathcal{O}_6^{hl}}{4m_Q^3}
 \,,
 \label{hqebvb}
\end{eqnarray}
where the matching coefficients $\tilde{C}_i$ are pure numbers. The coefficients of the four-quark operators
$\tilde C_i^{hl}$ are of $\mathcal{O}(\alpha_s^2)$ and therefore beyond the precision of the calculation, so they can be neglected.

The operator $\mathcal{O}_v$ in Eq.~(\ref{eq:HQE-1-dq}) can also be removed by using the equation of motion (EOM) of the HQET Lagrangian
\begin{eqnarray}
  \mathcal{O}_v &=&
 - \frac{1}{2m_Q} (\mathcal{O}_\pi+ c_F \mathcal{O}_G)
 -  \frac{1}{8m_Q^2} (c_D \mathcal{O}_D + c_S \mathcal{O}_{LS})
 \nonumber
 \\
 &&
 - \frac{1}{8 m_Q^2}(c_1^{hl} \mathcal{O}_1^{hl} + c_2^{hl} \mathcal{O}_2^{hl} + c_3^{hl} \mathcal{O}_3^{hl} + c_4^{hl} \mathcal{O}_4^{hl}  + c_5^{hl} \mathcal{O}_5^{hl}  + c_6^{hl} \mathcal{O}_6^{hl})
 \,,
 \label{LHQET}
\end{eqnarray}
where
\begin{eqnarray}
 c_F(\mu) &=& 1 + \frac{\alpha_s(\mu)}{2\pi}\bigg[ C_F + C_A\bigg(1 + \ln\left(\frac{\mu}{m_Q}\right)\bigg) \bigg]\,,
 \\
 c_D(\mu) &=& 1 + \frac{\alpha_s(\mu)}{\pi}\bigg[
 C_F\bigg(-\frac{8}{3}\ln\left(\frac{\mu}{m_Q}\right) \bigg)
 + C_A\bigg(\frac{1}{2} - \frac{2}{3}\ln\left(\frac{\mu}{m_Q}\right)\bigg) \bigg]\,,
\end{eqnarray}
are the coefficients of the chromomagnetic and Darwin operators in the HQET Lagrangian with NLO precision~\cite{Manohar:1997qy}. The coefficient of the spin-orbit operator $c_S = 2c_F -1$ is linked to the one of the
chromomagnetic operator due to reparametrization invariance~\cite{Luke:1992cs}. Again the coefficients $c_i^{hl}$ of the four-quark operators
are of $\mathcal{O}(\alpha_s^2)$ and therefore they can be neglected. The parameter $\mu$ is the renormalization scale and $C_F=4/3$, $C_A=3$ are color factors.

After all these considerations, the HQE for the inclusive semileptonic decay rate and $q^2$-spectrum is finally written as
\begin{eqnarray}
\Gamma(H_Q \rightarrow X_q \ell \bar \nu_\ell ) &=& \int_\eta^1 dr \frac{d\Gamma(H_Q \rightarrow X_q \ell \bar \nu_\ell )}{dr}
\nonumber
\\
  &=& \Gamma^0 |V_{qQ}|^2 \int_\eta^1 dr
 \bigg[ \mathcal{C}_0 \bigg( 1
- \frac{\bar{\mathcal{C}}_\pi - \bar{\mathcal{C}}_v }{\mathcal{C}_0}\frac{\mu_\pi^2}{2m_Q^2}\bigg)
+ \bigg(\frac{\bar{\mathcal{C}}_G}{c_F} -  \bar{\mathcal{C}}_v \bigg)\frac{\mu_G^2}{2m_Q^2}
\nonumber
\\
&&
- \bigg(\frac{\bar{\mathcal{C}}_D}{c_D}-\frac{1}{2} \bar{\mathcal{C}}_v \bigg) \frac{\rho_D^3}{2m_Q^3}
 - \bigg(\frac{\bar{\mathcal{C}}_{LS} }{c_S} - \frac{1}{2} \bar{\mathcal{C}}_v \bigg) \frac{\rho_{LS}^3}{2m_Q^3}
\nonumber
\\
&&
 	+ \mathcal{C}_1^{hl} \frac{\rho_1^{hl\,3}}{2m_Q^3}
	+ \mathcal{C}_2^{hl} \frac{\rho_2^{hl\,3}}{2m_Q^3}
	+ \mathcal{C}_3^{hl} \frac{\rho_3^{hl\,3}}{2m_Q^3}
	+ \mathcal{C}_4^{hl} \frac{\rho_4^{hl\,3}}{2m_Q^3}
    + \mathcal{C}_5^{hl} \frac{\rho_5^{hl\,3}}{2m_Q^3}
    + \mathcal{C}_6^{hl} \frac{\rho_6^{hl\,3}}{2m_Q^3}
 \bigg]
 \nonumber
 \\
 &\equiv& \Gamma^0 |V_{qQ}|^2 \int_\eta^1 dr
 \bigg(  \mathcal{C}_0
- \mathcal{C}_{\mu_\pi}\frac{\mu_\pi^2}{2m_Q^2}
+ \mathcal{C}_{\mu_G}\frac{\mu_G^2}{2m_Q^2}
- \mathcal{C}_{\rho_D} \frac{\rho_D^3}{2m_Q^3}
 - \mathcal{C}_{\rho_{LS}} \frac{\rho_{LS}^3}{2m_Q^3}
 \nonumber
\\
&&
 	+ \mathcal{C}_1^{hl} \frac{\rho_1^{hl\,3}}{2m_Q^3}
	+ \mathcal{C}_2^{hl} \frac{\rho_2^{hl\,3}}{2m_Q^3}
	+ \mathcal{C}_3^{hl} \frac{\rho_3^{hl\,3}}{2m_Q^3}
	+ \mathcal{C}_4^{hl} \frac{\rho_4^{hl\,3}}{2m_Q^3}
    + \mathcal{C}_5^{hl} \frac{\rho_5^{hl\,3}}{2m_Q^3}
    + \mathcal{C}_6^{hl} \frac{\rho_6^{hl\,3}}{2m_Q^3}
 \bigg)\,,
 \label{hqewidth2}
\end{eqnarray}
where the coefficients $\bar{\mathcal{C}}_i\equiv \mathcal{C}_i - \mathcal{C}_0 \tilde{C}_i$ are defined as the difference
between the coefficients $\mathcal{C}_i$ of the HQE of the transition operator in Eq.~(\ref{eq:HQE-1-dq}) and the current
in Eq.~(\ref{hqebvb}) multiplied by $\mathcal{C}_0$.
Note that reparametrization invariance also relates coefficients of higher dimensional operators to coefficients of lower dimensional operators in the HQE of the rate and $q^2$-spectrum, in particular $\mathcal{C}_0 = \mathcal{C}_{\mu_\pi}$ 
and $c_F \mathcal{C}_{\mu_G}=c_S \mathcal{C}_{\rho_{LS}}$~\cite{Manohar:2010sf,Becher:2007tk,Mannel:2018mqv,Fael:2018vsp}\footnote{Note that this relation was incorrectly extracted from~\cite{Manohar:2010sf} in
refs.~\cite{Mannel:2021zzr,Moreno:2022goo}. Now the relation is consistent with~\cite{Manohar:2010sf}.}.
The coefficients of the total rate $C_i$ ($i=0,\,\mu_\pi,\,\mu_G,\,\rho_D,\,\rho_{LS}$) are defined in analogy to Eq.~(\ref{Citot}).
The HQE hadronic parameters $\mu_\pi^2$, $\mu_G^2$, $\rho_D^3$, $\rho_{LS}^3$ and $\rho_i^{hl\,3}$ ($i=1,\ldots,6$) are defined as the following forward matrix elements of local HQET operators taken between full QCD states~\cite{Mannel:2018mqv}
\begin{eqnarray}
 \langle H_Q(p_{H_Q})\lvert \bar Q \slashed v Q \lvert H_Q(p_{H_Q})\rangle &=& 2M_{H_Q}\,, \label{eq:hadrforwardv} \\
 - \langle H_Q(p_{H_Q}) \lvert \mathcal{O}_\pi \lvert H_Q(p_{H_Q}) \rangle &=& 2M_{H_Q} \mu_\pi^2\,, \\
 c_F \langle H_Q(p_{H_Q}) \lvert \mathcal{O}_G \lvert H_Q(p_{H_Q}) \rangle
 &=& 2M_{H_Q} \mu_G^2\,, \\
  - c_D \langle H_Q(p_{H_Q})\lvert \mathcal{O}_D \lvert H_Q(p_{H_Q})\rangle&=& 4M_{H_Q} \rho_D^3\,, \\
 -  c_S \langle H_Q(p_{H_Q}) \lvert \mathcal{O}_{LS} \lvert H_Q(p_{H_Q})\rangle&=& 4 M_{H_Q} \rho_{LS}^3\,, \\
 \langle H_Q(p_{H_Q}) \lvert \mathcal{O}_i^{hl} \lvert H_Q(p_{H_Q}) \rangle&=& 4 M_{H_Q} \rho_i^{hl\,3}\,\quad\quad(i=1,\ldots,6).
 \label{ME4q}
\end{eqnarray}

\section{Computational Overview}
\label{sec:HQE}

In this section we address the computation of the coefficients of the Darwin and four-quark operators for the differential and total rates at NLO and obtain analytical results. The coefficients of
lower dimension operators are known and their expressions can be found within the definitions of this paper
in~\cite{Moreno:2022goo}.

For the computation we follow~\cite{Mannel:2021ubk,Mannel:2021zzr}. As a first step we compute the matching coefficients for the $q^2$-spectrum
by using the spectral representation given in Eq.~(\ref{spectrum1}) and matching to Eq.~(\ref{hqewidth2}). As a second step we
integrate over the dilepton invariant mass to obtain the coefficients of the
total rate.

Since Eqs. (\ref{eq:HQE-1}), (\ref{hqebvb}) and (\ref{LHQET}) hold at the operator level the matching coefficients can be computed by using quarks and gluons. In general, the matching is performed by comparing off shell amplitudes and latter using the EOM of HQET to remove operators that vanish on shell. For the leading power coefficient and the dimension six four-quark operators the matching is done by considering on shell heavy quarks. For power corrections the matching coefficients are computed by using a small momentum expansion near the heavy quark mass shell~\cite{Mannel:2015jka,Mannel:2019qel}.

The LO and NLO contributions to the Darwin coefficient of the differential rate are given by one-loop and two-loop heavy
quark to gluon-heavy quark ($Q\rightarrow g Q$) scattering amplitudes. The LO and NLO contributions to the four-quark operator coefficients of the differential rate are given by tree level and one-loop heavy-light quark to heavy-light quark ($Qq\rightarrow Qq$) scattering amplitudes.

We perform the calculation in Feynman gauge and use standard dimensional regularization with anticommuting $\gamma_5$ to treat both UV and IR divergences.
The scattering involving a gluon is computed in the external gluonic field by using the background field method. We use LiteRed~\cite{Lee:2012cn,Lee:2013mka} to reduce the corresponding amplitudes to a combination of the
master integrals given in~\cite{Mannel:2021ubk}.
Algebraic manipulations are carried out in Tracer~\cite{Jamin:1991dp}.

We use the $\overline{\mbox{MS}}$ renormalization scheme for the strong coupling $\alpha_s(\mu)$ and the HQET operators.
The heavy quark is renormalized on-shell
\begin{eqnarray}
 Q_B = (Z_2^{\mbox{\scriptsize OS}})^{1/2} Q\,,\quad\quad  m_{Q,B} = Z_{m_Q}^{\mbox{\scriptsize OS}} m_Q
 \,,\quad\quad 
 Z_2^{\mbox{\scriptsize OS}} &=& 1 - C_F \frac{\alpha_s(\mu)}{4\pi}\bigg( \frac{3}{\epsilon} + 6 \ln\left(\frac{\mu}{m_Q}\right) + 4 \bigg)\,,
\end{eqnarray}
where the subscript $B$ stand for bare quantities, and the ones without subscript stand for renormalized. To the order we are working on $Z_{m_Q}^{\mbox{\scriptsize OS}}=Z_2^{\mbox{\scriptsize OS}}$.
Therefore, the results for the coefficients are conveniently presented in the the on-shell scheme.

Note that in phenomenological applications one typically uses a short distance mass for the heavy quark in order to obtain more precise theoretical predictions, like
the $1S$ mass~\cite{Hoang:1999us,Hoang:1999zc,Hoang:1998hm,Hoang:1998ng,Bauer:2004ve} or the kinetic mass~\cite{Bigi:1994ga,Bigi:1996si}. Changing the mass scheme is possible by using the known one-loop relation between the different schemes.

After renormalization of the QCD couplings and fields there still remain IR divergences in the coefficient functions which point out the local operators in the effective theory develop UV divergences, or in other words, an anomalous dimension. These divergences cancel out after renormalization
of the corresponding operators in the HQE. In particular, operators of different dimension may mix under renormalization as it is the
case for the Darwin operator. The details on the renormalization of the effective operators including the discussion about the operator mixing is left to Secs.~\ref{sec:4qO} and \ref{sec:darwin}, where the computations of the four-quark operator coefficients and the Darwin operator coefficient are addressed, respectively.

For the presentation of results we split the Darwin coefficient of the differential and total rate in the following way
\begin{eqnarray}   
 \mathcal{C}_D(r,\eta) &=& \mathcal{C}_D^{\mbox{\scriptsize LO}}
 + \frac{\alpha_s}{\pi}\bigg(C_F \mathcal{C}_D^{\mbox{\scriptsize NLO, F}}
 + C_A \mathcal{C}_D^{\mbox{\scriptsize NLO, A}}\bigg)\,,
 \\
  C_D(\eta) &=& C_D^{\mbox{\scriptsize LO}}
 + \frac{\alpha_s}{\pi}\bigg(C_F C_D^{\mbox{\scriptsize NLO, F}}
 + C_A C_D^{\mbox{\scriptsize NLO, A}}\bigg)\,.
\end{eqnarray}
We provide analytical results for the Darwin and four-quark operator coefficients of the differential and total rate
in the ancillary file ``coefbutv.nb''. The differential rate is expressed in terms of delta and plus
distributions, where the latter is defined as
\begin{eqnarray}
 \int_\eta^1 dr f(r)\bigg[\frac{1}{1-r}\bigg]_{+} \equiv
 \int_\eta^1 dr \frac{f(r)-f(1)}{1-r} \,,
\end{eqnarray}
where $f(1)$ must be understood as expanded.

\subsection{Four-quark operator coefficients}
\label{sec:4qO}

Four-quark operators are the main responsible for lifetime differences between hadrons containing the same heavy quark but a different
spectator quark~\cite{Neubert:1996we,King:2021jsq}, as they explicitly involve the light quark field. In addition, they are phase space enhanced by a factor $16\pi^2$, 
overwhelming in general $SU(3)$ breaking effects in matrix elements of two-quark operators.

The coefficients of the dimension 6 four-quark operators for the inclusive semileptonic $H_Q \rightarrow X_q \ell \bar{\nu}_\ell$
decay rate are known up to NLO~\cite{Uraltsev:1996ta,Neubert:1996we,Beneke:2002rj,Franco:2002fc,Ciuchini:2001vx,Keum:1998fd,Lenz:2013aua}. To the best of my knowledge,
explicit expressions have never been presented to this order for the $q^2$-spectrum since previous studies are focused on their
effects to the lifetime differences. The four-quark operators may have a sizable impact due to their phase space enhancement. 
The computation of the four-quark operator coefficients up to NLO is also necessary for the computation of the Darwin
coefficient at NLO, since the corresponding operators mix under renormalization.

\begin{figure}[!htb]
	\centering
	\includegraphics[width=1.0\textwidth]{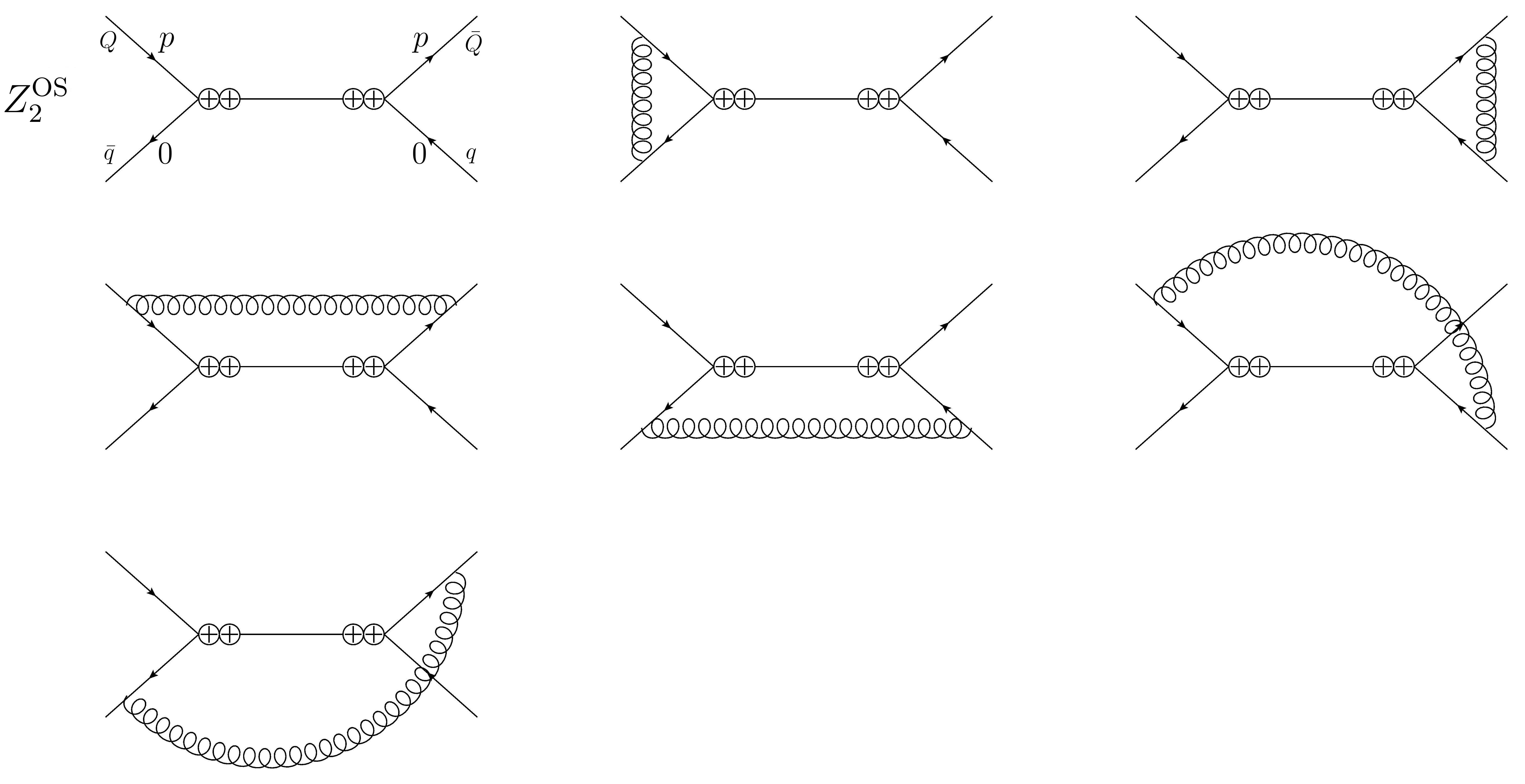}
        \caption{$Q(p)q(0) \rightarrow Q(p)q(0)$ scattering diagrams contributing to the LO and NLO four-quark operator coefficients
       $\mathcal{C}_i^{hl}$ ($i=1,\ldots,6$) in the HQE of the $H_Q \to X_q \ell \bar{\nu}_\ell$ decay spectrum, Eq.~(\ref{hqewidth2}).
        Circles with crosses stand for insertions of $\mathcal{L}_{\rm eff}$ and the thick line stands for the lepton-antineutrino
        propagator with mass $q$.}
        \label{match4q}
\end{figure}

\begin{figure}[!htb]
	\centering
	\includegraphics[width=0.9\textwidth]{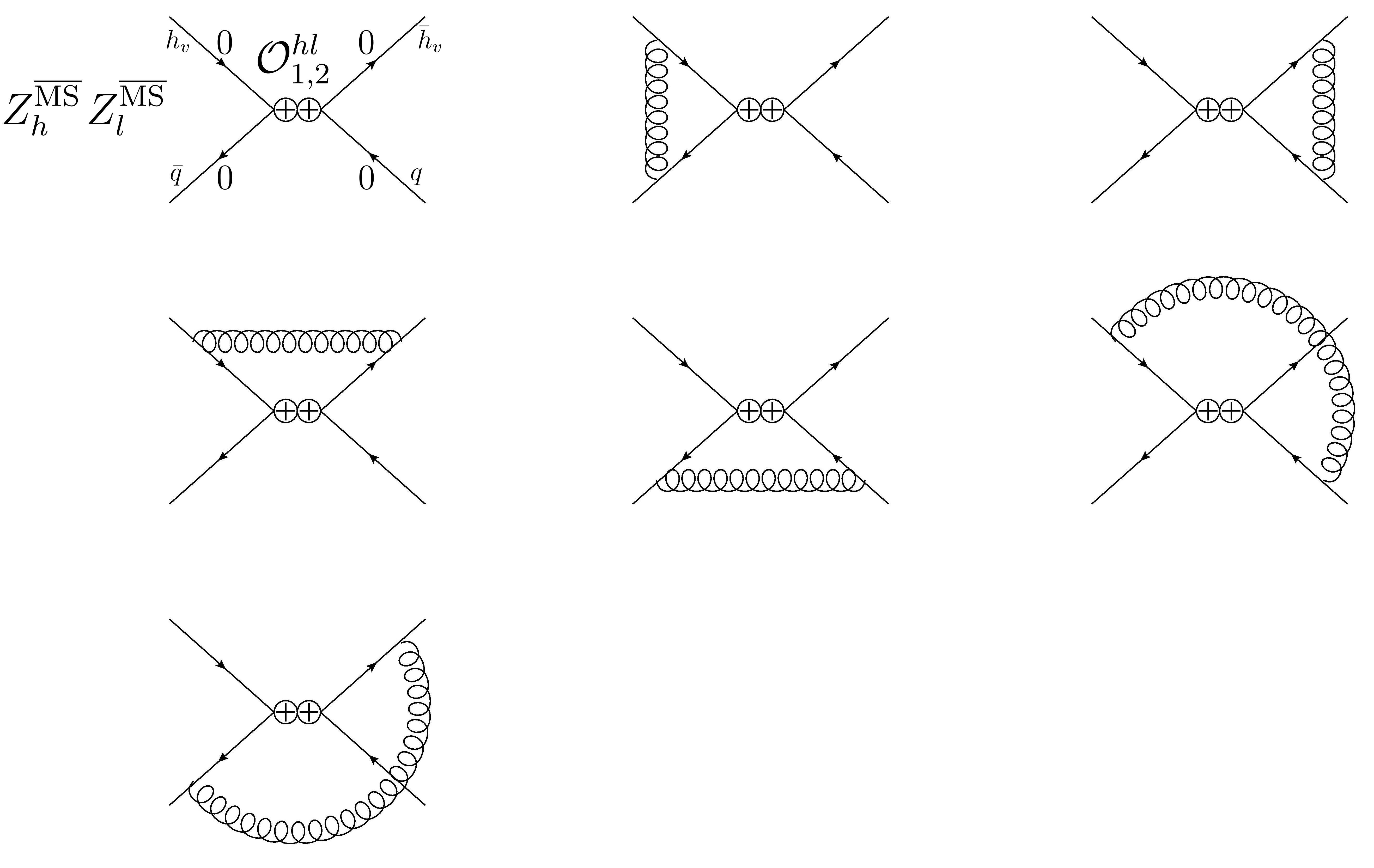}
        \caption{$h_v(0)q(0) \rightarrow h_v(0)q(0)$ scattering diagrams contributing to the operator mixing of four-quark operators
        with themselves at NLO. Circles with crosses stand for insertions of $\mathcal{O}_{1,2}^{hl}$.}
        \label{mixO4qO4q}
\end{figure}

For the computation, we take the $Q(p)q(k) \rightarrow Q(p)q(k)$ scattering amplitude with heavy quark momentum $p$, with
$p^2=m_Q^2$, and vanishing light quark momentum $k=0$. The diagrams that contribute to the coefficients of the differential rate
are shown in Fig.~\ref{match4q}. The first three diagrams only contribute to the singlet operators, whereas the remaining
diagrams contribute only to the octet operators.
The renormalized coefficients of the four-quark operators are then given by
\begin{eqnarray}
 \mathcal{C}_i^{hl} &=& \mathcal{C}_{i,\,B}^{hl} + \delta \mathcal{C}_i^{hl\,\rm \overline{MS}}
 \quad\quad (i=1,\ldots,6)\,,
 \label{CihlBCihlR}
\end{eqnarray}
where $\mathcal{C}_{i,\,B}^{hl}$ is defined as the sum of all diagrams of Fig.~[\ref{match4q}] including $Z_2$ and
$\delta \mathcal{C}_i^{hl\,\rm \overline{MS}}$ is the contribution due to the one-loop mixing under renormalization of the
effective operators, in particular, the mixing of four-quark operators with themselves. The coefficients
$\mathcal{C}_i^{hl}$ are finite and the cancellation of
poles provides a solid check of the calculation. To $\mathcal{O}(\alpha_s)$, the four-quark operators only mix with $\mathcal{O}_{1,2}^{hl}$ since these are the only operators whose coefficients get LO contributions.
The corresponding anomalous dimensions are obtained by computing the UV divergent part of the diagrams shown in Fig.~[\ref{mixO4qO4q}].
In this case the UV divergences can be computed by taking external heavy and light quarks with zero four-momenta and regulating IR divergences by using a gluon mass, which also sets an scale for the calculation. In Fig.~[\ref{mixO4qO4q}], the heavy and light wave function renormalization constants are
\begin{eqnarray}
 Z_h^{\rm \overline{MS}} = 1 + 2 C_F \frac{\alpha_s}{4\pi}\frac{1}{\epsilon}\,, \quad \quad
 Z_l^{\rm \overline{MS}} = 1 -  C_F \frac{\alpha_s}{4\pi}\frac{1}{\epsilon}\,,
\end{eqnarray}
respectively. We find the following counterterms due to operator mixing
\begin{eqnarray}
 \delta \mathcal{C}_1^{hl\,\rm \overline{MS}} &=& 3\mathcal{C}_1^{hl} C_F \frac{\alpha_s}{4\pi}\frac{1}{\epsilon},\,\quad\quad 
 \delta \mathcal{C}_2^{hl\,\rm \overline{MS}} = 3\mathcal{C}_2^{hl} C_F \frac{\alpha_s}{4\pi}\frac{1}{\epsilon},\,\quad\quad
 \delta \mathcal{C}_3^{hl\,\rm \overline{MS}} = \mathcal{C}_1^{hl} \frac{\alpha_s}{4\pi}\frac{1}{\epsilon},
 \nonumber
 \\
  \delta \mathcal{C}_4^{hl\,\rm \overline{MS}} &=& -2\mathcal{C}_2^{hl} \frac{\alpha_s}{4\pi}\frac{1}{\epsilon},\,\quad\quad
 \delta \mathcal{C}_5^{hl\,\rm \overline{MS}} = - \frac{1}{4}\mathcal{C}_2^{hl} \frac{\alpha_s}{4\pi}\frac{1}{\epsilon},\,\quad\quad
 \delta \mathcal{C}_6^{hl\,\rm \overline{MS}} = - \frac{1}{4}\mathcal{C}_1^{hl} \frac{\alpha_s}{4\pi}\frac{1}{\epsilon}.
\end{eqnarray}
In the massive lepton case, the four-quark operator coefficients of the differential rate read
\begin{eqnarray}
 \mathcal{C}_1^{hl} &=&
      -256\pi^2 (1 - \eta)^2 (2 + \eta)
      \bigg[
       1 - C_F \frac{\alpha_s}{\pi} \bigg(2 + \frac{3}{2} \ln\left(\frac{\mu}{m_Q}\right)
      \bigg)
      \bigg]
      \delta(-1 + r) \,,
\\
 \mathcal{C}_2^{hl} &=&
      256 \pi^2 (1 - \eta)^2 \bigg[
       2 (1 + 2 \eta) -
   C_F \frac{\alpha_s}{\pi} \bigg( 4 + 5 \eta + 3(1 + 2\eta) \ln\left(\frac{\mu}{m_Q}\right) \bigg)
      \bigg]
      \delta(-1 + r)\,,
\\
\mathcal{C}_3^{hl} &=&
  - \frac{\alpha_s}{\pi} 64 \pi^2 \bigg\{
    \frac{2}{3 r^3}  (r - \eta)^2 \bigg[
        2\eta
       + r(1 + 9 \eta)
       + 3r^2 (5 - 2 \eta)
       - 2r^3 (3 + \eta)
       - 4r^4
      \bigg]
      \bigg[\frac{1}{1-r}\bigg]_{+}
      \nonumber
      \\
      &&
      +  (1 - \eta)^2 \bigg[
       15 + 9 \eta + 2 (2 + \eta) \ln(1 - \eta) - 2 (2 + \eta) \ln\left(\frac{\mu}{m_Q}\right)
      \bigg]
      \delta(-1 + r)
      \bigg\}\,,
      \\
      \mathcal{C}_4^{hl} &=&
    \frac{\alpha_s}{\pi} 128 \pi^2\bigg\{
     \frac{2}{3 r^3} (r - \eta)^2 \bigg[
      2\eta + r(1 - 8\eta) - r^2 (1 + 8 \eta) - 2r^3 (5 - \eta) + 4r^4
      \bigg]
      \bigg[\frac{1}{1-r}\bigg]_{+}
      \nonumber
      \\
      &&
      + (1 - \eta)^2 \bigg[
       3 (3 + 4 \eta) - 4(1 + 2\eta) \ln(1 - \eta) + 4 (1 + 2 \eta) \ln\left(\frac{\mu}{m_Q}\right)
      \bigg]
      \delta(-1 + r)
      \bigg\}\,,
      \\
      \mathcal{C}_5^{hl} &=&
   - \frac{\alpha_s}{\pi} 32 \pi^2\bigg\{
    \frac{2}{3 r^3} (r - \eta)^2  \bigg[
      10\eta
     + r(5 - \eta)
     + r^2 (1 - \eta)
     + r^3 (1 - 2 \eta)
     - 4r^4
      \bigg]
      \bigg[\frac{1}{1-r}\bigg]_{+}
      \nonumber
      \\
      &&
      + (1 - \eta)^2 \bigg[
       2 + \eta + 2(1 + 2\eta) \ln(1 - \eta) - 2 (1 + 2 \eta) \ln\left(\frac{\mu}{m_Q}\right)
      \bigg]
      \delta(-1 + r)
      \bigg\}\,,
      \\
      \mathcal{C}_6^{hl} &=&
     \frac{\alpha_s}{\pi} 32 \pi^2 \bigg\{
       \frac{1}{3 r^2} (r - \eta)^2
       (5 - r - r^2) (\eta + 2r)
      \bigg[\frac{1}{1-r}\bigg]_{+}
      \nonumber
      \\
      &&
      + (1 - \eta)^2 (2 + \eta) \bigg[
       \ln(1 - \eta) - \ln\left(\frac{\mu}{m_Q}\right)
      \bigg]
      \delta(-1 + r)
      \bigg\}\,,
\end{eqnarray}
whereas in the massless lepton case ($\eta=0$), they read
\begin{eqnarray}
 \mathcal{C}_1^{hl} &=&
      - 512\pi^2 \bigg[
       1  - C_F \frac{\alpha_s}{\pi} \bigg( 2 + \frac{3}{2} \ln\left(\frac{\mu}{m_Q}\bigg) \right)
      \bigg]
      \delta(-1 + r)\,,
      \\
      \mathcal{C}_2^{hl} &=&
      512 \pi^2 \bigg[
       1 - C_F \frac{\alpha_s}{\pi} \bigg(2 + \frac{3}{2} \ln\left(\frac{\mu}{m_Q}\right) \bigg)
      \bigg]
      \delta(-1 + r)\,,
      \\
      \mathcal{C}_3^{hl} &=&
    - \frac{\alpha_s}{\pi}64 \pi^2
    \bigg\{
    \frac{2}{3}(1 + 15 r - 6 r^2 - 4 r^3)
      \bigg[\frac{1}{1-r}\bigg]_{+}
      + \bigg[ 15 - 4 \ln\left(\frac{\mu}{m_Q}\right)
      \bigg]
      \delta(-1 + r)
      \bigg\}\,,
      \\
      \mathcal{C}_4^{hl} &=&
   \frac{\alpha_s}{\pi}  128 \pi^2 \bigg\{
     \frac{2}{3}
    (1 - r - 10 r^2 + 4 r^3)
      \bigg[\frac{1}{1-r}\bigg]_{+}
      +  \bigg[
       9 + 4 \ln\left(\frac{\mu}{m_Q}\right)
      \bigg]
      \delta(-1 + r)
      \bigg\}\,,
      \\
 \mathcal{C}_5^{hl} &=&
      -\frac{\alpha_s}{\pi}64\pi^2 \bigg\{
      \frac{1}{3} (5 + r + r^2 - 4 r^3)
      \bigg[\frac{1}{1-r}\bigg]_{+}
      + \bigg[
       1 - \ln\left(\frac{\mu}{m_Q}\right)
      \bigg]
      \delta(-1 + r)
      \bigg\}\,,
      \\
      \mathcal{C}_6^{hl} &=&
     \frac{\alpha_s}{\pi} 64 \pi^2 \bigg\{
       \frac{1}{3}  r (5 - r - r^2)
      \bigg[\frac{1}{1-r}\bigg]_{+}
      - \ln\left(\frac{\mu}{m_Q}\right)
      \delta(-1 + r)
      \bigg\}\,.
\end{eqnarray}
The coefficients of the total rate are obtained from Eq.~(\ref{Citot}) by integrating the coefficients of the differential rate
over $r$ in the whole range. In the massive lepton case we obtain
\begin{eqnarray}
 C_1^{hl} &=&
 -256 \pi^2 (\eta-1)^2 (\eta + 2) \bigg[
  1
 - C_F \frac{\alpha_s}{\pi} \left(2 + \frac{3}{2} \ln \left(\frac{\mu }{m_Q}\right)\right)
 \bigg]\,,
 \\
 C_2^{hl} &=&
 256\pi ^2(\eta -1)^2 \bigg[
 2(2 \eta +1)
 - C_F \frac{\alpha_s}{\pi}  \left(5 \eta + 4 + 3(2 \eta +1) \ln \left(\frac{\mu }{m_Q}\right)\right)
 \bigg]\,,
 \\
 C_3^{hl} &=&
 -\frac{\alpha_s}{\pi} \frac{64}{9} \pi ^2 \bigg[
  (\eta -1) (19\eta^2 + 118\eta - 143)
 - 6(5 \eta -6) \eta^2 \ln (\eta )
 \nonumber
 \\
 &&
 + 18 (\eta-1)^2 (\eta + 2) \bigg( \ln (1-\eta ) - \ln \left(\frac{\mu }{m_Q}\right)\bigg)
   \bigg]\,,
   \\
   C_4^{hl} &=&
 \frac{\alpha_s}{\pi}\frac{128}{9} \pi ^2 \bigg[
  (\eta -1)(128 \eta^2 -7\eta-133)
 + 12 \eta ^2 (7 \eta -6) \ln (\eta )
  \nonumber
 \\
 &&
  - 36 (2 \eta +1) (\eta -1)^2 \bigg( \ln (1-\eta ) - \ln \left(\frac{\mu }{m_Q}\right)\bigg)
   \bigg]\,,
   \\
   C_5^{hl} &=&
 \frac{\alpha_s}{\pi}\frac{32}{9} \pi ^2 \bigg[
   (\eta -1)(65\eta^2 -88\eta+47)
 + 24 \eta ^2 (2 \eta -3) \ln (\eta )
 \nonumber
 \\
 &&
 - 18 (2 \eta +1) (\eta -1)^2 \bigg( \ln (1-\eta ) - \ln \left(\frac{\mu }{m_Q}\right)\bigg)
   \bigg]\,,
   \\
   C_6^{hl} &=&
 -\frac{\alpha_s}{\pi}\frac{16}{9} \pi ^2 \bigg[
 25 \eta^3 - 54\eta^2 + 9\eta + 20
+ 24 \eta ^3 \ln (\eta )
 \nonumber
 \\
 &&
 - 18 (\eta-1)^2 (\eta + 2) \bigg( \ln (1-\eta ) - \ln \left(\frac{\mu }{m_Q}\right) \bigg)
   \bigg]\,.
\end{eqnarray}
Finally, in the massless lepton ($\eta=0$) case we obtain
\begin{eqnarray}
 C_1^{hl} &=& - 512\pi^2 \bigg[
 1 - C_F \frac{\alpha_s}{\pi} \left(2 + \frac{3}{2} \ln \left(\frac{\mu }{m_Q}\right) \right)
 \bigg]\,,
 \\
 C_2^{hl} &=&
 512\pi^2 \bigg[
 1 - C_F \frac{\alpha_s}{\pi}  \left(2 + \frac{3}{2} \ln \left(\frac{\mu }{m_Q}\right)\right)
 \bigg]\,,
 \\
 C_3^{hl} &=& -\frac{\alpha_s}{\pi}\frac{64}{9} \pi ^2 \left(143 - 36 \ln \left(\frac{\mu }{m_Q}\right)\right)\,,
 \\
 C_4^{hl} &=& \frac{\alpha_s}{\pi}\frac{128}{9} \pi ^2 \left(133 + 36 \ln \left(\frac{\mu }{m_Q}\right)\right)\,,
 \\
 C_5^{hl} &=& -\frac{\alpha_s}{\pi}\frac{32}{9} \pi ^2 \left(47 - 18 \ln \left(\frac{\mu }{m_Q}\right)\right)\,,
 \\
 C_6^{hl} &=& -\frac{\alpha_s}{\pi}\frac{64}{9} \pi ^2 \left(5 + 9 \ln \left(\frac{\mu }{m_Q}\right) \right)\,.
\end{eqnarray}
Note that in the massless lepton case the singlet operators combine in perpendicular form
$\mathcal{O}_\perp^{hl} \equiv \mathcal{O}_1^{hl} - \mathcal{O}_2^{hl}$. Matrix elements of such an operator are
zero in the vacuum insertion approximation (VIA). Likewise, in such an approximation the matrix elements
of octet operators over singlet states are also zero. Even though the computation of these matrix elements by using HQET
sum rules~\cite{Kirk:2017juj,King:2021jsq} show deviations from VIA, it works well as a first approximation~\cite{Lenz:2022rbq}. Therefore, in this particular case the phase space enhancement can be
easily canceled by the small value of the matrix elements themselves. It is quite remarkable that in the massive lepton
case the singlet operators do not combine in $\mathcal{O}_\perp^{hl}$, and therefore their matrix elements are not suppressed.
Therefore, the four-quark operators will have a more visible impact in semitauonic decays than in its electron or muon counterparts.

After changing the operator basis to the evanescent operator basis used in~\cite{Buras:1989xd,Beneke:2002rj} the NLO results for the total width agree with~\cite{Beneke:2002rj,Lenz:2013aua}. The change in the operator basis and the corresponding comparison is discussed in more detail in Sec.~\ref{sec:ev}. Note that in~\cite{Beneke:2002rj,Lenz:2013aua} the coefficients were computed by directly using an operator basis with evanescent operators whereas we have chosen a basis with arbitrarily long strings of gamma matrices and later we have done the change of basis. That constitutes a strong check of the calculation and confirms the former results.

\subsection{Darwin operator coefficient}
\label{sec:darwin}
The Darwin coefficient of the total rate in inclusive semileptonic decays with a massless final state quark
can be extracted to LO from the corresponding calculation in non-leptonic
decays~\cite{Lenz:2020oce,Mannel:2020fts,Moreno:2020rmk}. To the best of my knowledge it has never been computed for the
differential rate to this order. We address the computation of the Darwin coefficient at NLO for both the total rate and the
$q^2$-spectrum, which has never been considered before in the literature.

\begin{figure}[!htb]
	\centering
	\includegraphics[width=1.0\textwidth]{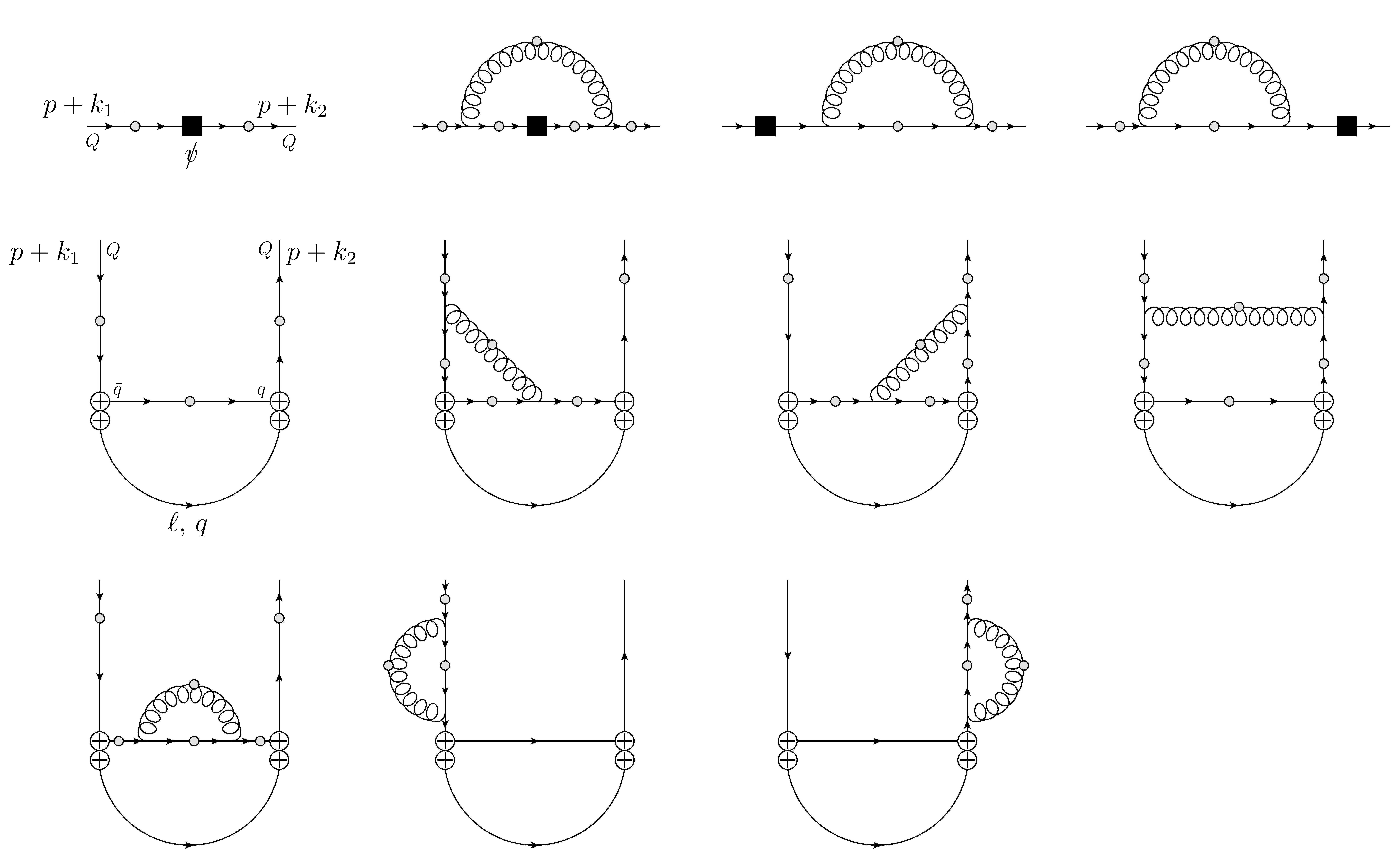}
        \caption{$Q(p+k_1)\rightarrow Q(p+k_2)g(k_1-k_2)$ scattering diagrams contributing to the LO and NLO coefficients
        $\bar{\mathcal{C}}_i$ of power corrections in the HQE of the $H_Q \to X_q \ell \bar{\nu}_\ell$ decay spectrum, Eq.~(\ref{hqewidth2}). Black squares stand for $\slashed v$ insertions,
        circles with crosses for insertions of $\mathcal{L}_{\rm eff}$, thick lines for the lepton-antineutrino propagator with mass $q$, and gray dots for possible one gluon insertions
        with outgoing momentum $k_1-k_2$. After accounting for all one gluon insertions, there are five diagrams at LO and
        forty-one diagrams at NLO.}
        \label{SampleFDdifwidth}
\end{figure}

For the the computation we follow~\cite{Mannel:2021zzr,Moreno:2022goo} and take the $Q(p+k_1)\rightarrow Q(p+k_2)g(k_1-k_2)$
scattering amplitude $\mathcal{A}$ in QCD with hard momentum $p$, with $p^2 = m_Q^2$, and two soft momenta $k_1$, $k_2$, and
expand it to quadratic order in the small momenta (the loop momenta are hard)
\begin{eqnarray}
 \mathcal{A} = I_0
 + I_{k_1}^\alpha {k_1}_\alpha
 + I_{k_2}^\alpha {k_2}_\alpha
 + I_{k_1^2}^{\alpha\beta} {k_1}_\alpha {k_1}_\beta
 + I_{k_2^2}^{\alpha\beta} {k_2}_\alpha {k_2}_\beta
 + I_{k_1 k_2}^{\alpha\beta} {k_1}_\alpha {k_2}_\beta + \ldots\,,
 \label{ampk1k2}
\end{eqnarray}
where dots stand for higher orders in the expansion. The diagrams that contribute are shown in Fig.~[\ref{SampleFDdifwidth}].
The coefficient of the Darwin operator is obtained
by projecting the amplitude to the corresponding operator. This is achieved by taking
the contribution proportional to ${k_1}_\alpha {k_2}_\beta$ in Eq.~(\ref{ampk1k2}) and using the projector
$\mathcal{P}_{\mathcal{O}_D} = (g_{\alpha\beta} - v_\alpha v_\beta ) P_{+}$ such that
\begin{eqnarray}
 \mathcal{C}_D  &\sim& \Tr[ v_\lambda (g_{\alpha\beta} - v_\alpha v_\beta ) I_{k_1 k_2}^{\lambda\alpha\beta} P_{+}]\,,
\end{eqnarray}
where $P_{+} = (1 + \slashed v)/2$. This projection ensures that the contribution to the Darwin coefficient is disentangled
from the spin-orbit operator and operators which contribute to higher orders in the power expansion after using the EOM, but which
nevertheless merge with the Darwin operator before using the EOM (see e.~g.~\cite{Mannel:2021zzr} for more details).

In practice we directly compute the difference between the HQE of the transition operator and the current
\begin{eqnarray}
\bar{\mathcal{C}}_D &\equiv& \mathcal{C}_D - \mathcal{C}_0  \tilde{C}_D
\nonumber
\\
&=& Z_2^{\mbox{\scriptsize OS}}Z_{\mathcal{O}_D}(\mathcal{C}_{D,B} - \mathcal{C}_{0,B}  \tilde{\mathcal{C}}_{D,B})
 + \delta \mathcal{C}_D^{\rm \overline{MS},\, 2q\,(1)}
 + \delta \mathcal{C}_D^{\rm \overline{MS},\,4q\,(1)}
 + \delta \mathcal{C}_D^{\rm \overline{MS},\,4q\,(2)}\,,
\end{eqnarray}
where
\begin{eqnarray}
 Z_{\mathcal{O}_D} &=& - \frac{1}{6}C_A \frac{\alpha_s}{\pi}\frac{1}{\epsilon}\,,
 \label{CDmix1}
 \\
\delta \mathcal{C}_D^{\rm \overline{MS},\, 2q\,(1)} &=&
  \bigg[
   C_F\bigg(
     \frac{4}{3}\bar{\mathcal{C}}_{\pi,\,B}
   - \frac{2}{3}\bar{\mathcal{C}}_{v,\,B}
   \bigg)
 + C_A\bigg(
  \frac{5}{12}\bar{\mathcal{C}}_{G,\,B}
 + \frac{1}{12}\bar{\mathcal{C}}_{\pi,\,B}
 - \frac{1}{4}\bar{\mathcal{C}}_{v,\,B}
 \bigg)
 \bigg]
 \frac{\alpha_s}{\pi}\frac{1}{\epsilon}\,,
 \label{CDmix2}
 \\
 \delta \mathcal{C}_D^{\rm \overline{MS},\,4q\,(1)} &=&
 \bigg[ 2\mathcal{C}_{1,\,B}^{hl} - \mathcal{C}_{2,\,B}^{hl}
 \nonumber
 \\
 &&
 + (2\mathcal{C}_{3,\,B}^{hl} - \mathcal{C}_{4,\,B}^{hl} - 4\mathcal{C}_{5,\,B}^{hl} + 32\mathcal{C}_{6,\,B}^{hl})\left(C_F - \frac{C_A}{2}\right)\bigg]
 \frac{-1}{96\pi^2\epsilon}\bar{\mu}^{-2\epsilon}\,,
 \label{CDmix3}
 \\
 \delta \mathcal{C}_D^{\rm \overline{MS},\,4q\,(2)} &=&
  \bigg[
 \frac{5}{6}C_A (2\mathcal{C}_{1,\,B}^{hl} - \mathcal{C}_{2,\,B}^{hl})\frac{1}{\epsilon^2}
 \nonumber
 \\
 &&
 + \bigg(
  C_F( 2\mathcal{C}_{1,\,B}^{hl} - 5 \mathcal{C}_{2,\,B}^{hl})
 + \frac{5}{6} C_A\left( \frac{31}{3} \mathcal{C}_{1,\,B}^{hl} - \frac{1}{6}\mathcal{C}_{2,\,B}^{hl}\right)
 \bigg)\frac{1}{\epsilon}
 \bigg]\frac{\alpha_s}{4\pi}\frac{-1}{192\pi^2}\bar{\mu}^{-2\epsilon}\,,
 \label{CDmix4}
\end{eqnarray}
are the renormalization factor of the Darwin operator and the contributions to the Darwin coefficient due to operator mixing under renormalization given by the one-loop mixing of the two-quark operators, 
the one-loop mixing of the four-quark operators and the two-loop mixing of the four-quark operators with the Darwin operator, respectively. 
The quantity $\bar{\mathcal{C}}_D$ is finite and the cancellation of
poles provides a solid check of the calculation. Finally $ \bar{\mu}^{-2\epsilon} = \mu^{-2\epsilon}(e^{\gamma_E}/(4\pi))^{-\epsilon}$
is the $\overline{\mbox{MS}}$ renormalization scale.

The one-loop operator mixing due to two-quark operators is
known~\cite{Falk:1990pz,Bauer:1997gs,Finkemeier:1996uu,Balzereit:1996yy,Blok:1996iz,Lee:1991hp,Moreno:2017sgd,Lobregat:2018tmn,Moreno:2018lbo}. The quantities $Z_{\mathcal{O}_D}$ and $\delta C_D^{\rm \overline{MS},\, 2q\,(1)}$ are taken
from~\cite{Mannel:2021zzr,Moreno:2022goo}. The one-loop operator mixing due to four-quark operators can be extracted
from~\cite{Mannel:2020fts,Moreno:2020rmk}. Note that the first line of Eq.~(\ref{CDmix3}) contributes to the
renormalization of Darwin coefficient at LO and NLO when the coefficients $\mathcal{C}_{1-2\,,B}^{hl}$ are taken at LO and NLO, respectively. 
Indeed this is the only contribution due to operator mixing needed to renormalize the Darwin coefficient at LO. 
The bare coefficients $\mathcal{C}_{i\,,B}^{hl}$ are defined as in Eq.~(\ref{CihlBCihlR}). Note that
the coefficients $\mathcal{C}_{3-6,\,B}^{hl}$ contribute to the renormalization of the Darwin operator at NLO only through the one-loop operator mixing due to the corresponding coefficients are of $\mathcal{O}(\alpha_s)$.
The two-loop operator mixing due to four-quark operators is not known and the corresponding contribution to the Darwin coefficient is
inferred from the cancellation of the poles, which is achieved for a single combination of the coefficients
$\mathcal{C}_{1-2}^{hl}$, which are the only coefficients that are non-zero at LO, and therefore the only ones that contribute at
two-loops. This cancellation is not straightforward due to the non-trivial dependence of the coefficients $\mathcal{C}_{1-2}^{hl}$ in
$\eta$ and $r$. We find there is actually a unique combination that cancels the poles, which provides a check of the calculation.
Finally, note that the structure of the $1/\epsilon^2$ in pole in Eq.~(\ref{CDmix4}) is the same that appears in the $1/\epsilon$ pole in the first line of in Eq.~(\ref{CDmix3}), 
i.e. the corresponding poles are proportional to $2\mathcal{C}_{1,\,B}^{hl} - \mathcal{C}_{2,\,B}^{hl}$. 
This is to be expected due to the link between the $1/\epsilon$ and $1/\epsilon^2$ poles in the one-loop and two-loop anomalous dimensions and provides
an additional check of the computation.

Renormalization can be performed for the differential rate, as it holds at the level of the hadronic tensor. However,
it is more involved than for the case of a massive final-state quark due to the operator mixing with four-quark operators.
The cancellation of poles at the differential level requires the use of plus distributions due to the apparently different functional
structure of the contribution coming from $Q \rightarrow Qg$ scattering diagrams and the contribution coming from the operator mixing with four-quark operators. 
Whereas the former gives contributions proportional to $(1-r)^{-1-n\epsilon}$ which generate poles only after integration over $r$, the latter gives singular ($1/\epsilon$) contributions proportional 
to $\delta(-1+r)$. The cancellation of poles is achieved after using
\begin{eqnarray}
 \int_\eta^1 dr \frac{f(r,\epsilon)}{(1-r)^{1+n\epsilon}} \equiv
 \int_\eta^1 dr f(r,\epsilon)(1-r)^{-n\epsilon}\bigg[\frac{1}{1-r}\bigg]_{+}
 + \int_\eta^1 dr\frac{-1}{n\epsilon}(1-\eta)^{-n\epsilon}  f(1,\epsilon) \delta(-1+r)\,,
\end{eqnarray}
where the singular delta term which is generated produces the required contribution to cancel the one coming from the
operator mixing. Note that the right hand side of the equation above can be safely expanded in $\epsilon$ before integration.
Also note the close connection between dimensionally regularized IR singular integrals at the endpoint and delta functions sitting at that endpoint.

Finally, the Darwin coefficient of the differential rate $\mathcal{C}_{\rho_D}$ is obtained from
\begin{equation}
 \mathcal{C}_{\rho_D} = \frac{\bar{\mathcal{C}}_D}{c_D(\mu)} - \frac{1}{2} \bar{\mathcal{C}}_v\,.
\end{equation}
In the massive lepton case, the Darwin operator coefficient of the differential rate reads
\begin{eqnarray}
 \mathcal{C}_{\rho_D}^{\rm LO} &=& \frac{2}{3r^3} (r - \eta)^2
 (11 - 9 r + 9 r^2 + 5 r^3)  ( 2 \eta + r(1 + \eta) + 2 r^2 )
      \bigg[\frac{1}{1-r}\bigg]_{+}
      \nonumber
      \\
      &&
      + \frac{16}{3} (1 - \eta)^2 \bigg[
      5 + 4 \eta + 6 (1 + \eta) \ln(1 - \eta)
      - 6 (1 + \eta) \ln\left(\frac{\mu}{m_Q}\right)
      \bigg] \delta(-1 + r)\,,
      \\
      &&
 \nonumber
 \\
 \mathcal{C}_{\rho_D}^{\rm NLO,\,F} &=&
 \frac{1}{18 r^5} (r - \eta)^2 \bigg[
  64r\eta - 8r^2 (2 - 73 \eta) + r^3 (217 - 797 \eta) - 11 r^4 (71 + 7 \eta)
  \nonumber
  \\
  &&
  - 7 r^5 (25 - 39 \eta) + r^6 (273 + \eta) + 2r^7
  \nonumber
  \\
  &&
  + 2 \Big(
   32 \eta
  - r (8 + 5 \eta)
  - r^2 (28 + 487 \eta)
  - r^3 (221 - 101 \eta)
  \nonumber
  \\
  &&
  - r^4 (293 + 547 \eta)
  - r^5 (419 - 4 \eta)
  + r^6 (29 + 38 \eta)
  + 76 r^7
  \Big) \ln(1 - r)
  \nonumber
  \\
  &&
  - 4 r^3 \Big( 42\eta + 3r (7 - 36 \eta) - r^2 (108 - 59 \eta) + r^3 (61 + 19 \eta) + 38r^4
    \Big) \ln(r)
        \nonumber
  \\
  &&
      + 32r^2 \Big( 22\eta
      + r(11 + 26 \eta)
      + r^2 (43 + 9 \eta)
      + r^3 (9 + 10 \eta)
      - 4r^4 (1 - 2\eta)
            \nonumber
  \\
  &&
       + 16r^5
      \Big) \ln\left(\frac{\mu}{m_Q}\right)
        \nonumber
  \\
  &&
      - 6 (1 - r) r^2 \Big(
       14\eta
      + r(7 - 29 \eta)
      - 4r^2 (7 + 5\eta)
      - 5r^3 (5 + \eta)
      - 10r^4
      \Big)
       \nonumber
      \\
      &&
      \;\;\;\;\times
      \Big(\pi^2 - 2 \Li_2(1 - r) + 2 \Li_2(r) \Big)
 \bigg]
 \bigg[\frac{1}{1-r}\bigg]_{+}
  \nonumber
  \\
  &&
 - \frac{1}{9} (1 - \eta)^2 \bigg[
  159 + 81\eta + 88\pi^2 (1 + \eta)
  + 24 (10 - \eta) \ln(1 - \eta)
    \nonumber
  \\
  &&
  + 432 (1 + \eta) \ln^2(1 - \eta)
  - 40 (13 + 5 \eta) \ln\left(\frac{\mu}{m_Q}\right)
    \nonumber
  \\
  &&
  - 1200 (1 + \eta) \ln(1 - \eta) \ln\left(\frac{\mu}{m_Q}\right)
  + 768 (1 + \eta) \ln^2\left(\frac{\mu}{m_Q}\right)
 \bigg]
   \delta(-1 + r) \,,
\end{eqnarray}
\begin{eqnarray}
 \mathcal{C}_{\rho_D}^{\rm NLO,\,A} &=&
 -\frac{1}{27 r^5} (r - \eta)^2 \bigg[
  48r \eta
  - 4r^2 (3 - 38 \eta)
  + r^3 (67 + 91 \eta)
  + 2r^4 (91 + 72 \eta)
  \nonumber
  \\
  &&
      - r^5 (117 - 431 \eta)
      + 2 r^6 (209 + 74 \eta)
      + 296r^7
  \nonumber
  \\
  &&
      + 6 \Big(
        8\eta
      - r (2 + 17 \eta)
      - r^2 (7 + 20 \eta)
      - r^3 (28 + 39 \eta)
      + r^4 (21 - 46 \eta)
       \nonumber
 \\
 &&
      - 10r^5 (11 - 4 \eta)
      + 2r^6 (19 + 7 \eta)
      + 28 r^7
      \Big) \ln(1 - r)
       \nonumber
 \\
 &&
      + 6r^3 \Big(
       6\eta
      + 3r
      - r^2 (18 + 7 \eta)
      + 14r^3 (2 - \eta)
      - 28r^4
      \Big) \ln(r)
       \nonumber
 \\
 &&
      - 36r^2 \Big(
       4\eta
      + 2r(1 - 4\eta)
      - 2r^2 (5 + 3 \eta)
      + r^3 (9 + 7 \eta)
      - r^4 (13 - 9 \eta)
       \nonumber
 \\
 &&
      + 18r^5
      \Big) \ln\left(\frac{\mu}{m_Q}\right)
       \nonumber
 \\
 &&
      + 9 (1 - r) r^2 \Big( 2\eta + r(1 + \eta) + 2r^2 (1 + 2\eta) + 8r^3 \Big)
       \nonumber
 \\
 &&
 \;\;\;\;\times
      \Big(\pi^2 - 2 \Li_2(1 - r) + 2 \Li_2(r) \Big)
 \bigg]
 \bigg[\frac{1}{1-r}\bigg]_{+}
 \nonumber
 \\
 &&
 - \frac{1}{108} (1 - \eta)^2 \bigg[
   5381 + 4743\eta
  - 360 \pi^2 (1 + \eta)
  + 24(139 + 169 \eta) \ln(1 - \eta)
    \nonumber
 \\
 &&
   - 720 (1 + \eta) \ln^2(1 - \eta)
   - 24 (219 + 233 \eta) \ln\left(\frac{\mu}{m_Q}\right)
   \nonumber
 \\
 &&
  - 864 (1 + \eta) \ln(1 - \eta) \ln\left(\frac{\mu}{m_Q}\right)
  + 1584 (1 + \eta) \ln^2\left(\frac{\mu}{m_Q}\right)
 \bigg]
   \delta(-1 + r)\,,
\end{eqnarray}
where $\Li_2(x)$ is the dilogarithm. In the massless lepton case ($\eta=0$) it reads
\begin{eqnarray}
 \mathcal{C}_{\rho_D}^{\rm LO} &=&
      \frac{2}{3} (11 + 13 r - 9 r^2 + 23 r^3 + 10 r^4)
      \bigg[\frac{1}{1-r}\bigg]_{+}
      + \bigg[
      \frac{80}{3} - 32 \left(\frac{\mu}{m_Q}\right)
      \bigg]
      \delta(-1 + r)\,,
      \nonumber
      \\
      &&
 \\
 \mathcal{C}_{\rho_D}^{\rm NLO,\,F} &=&
      -\frac{1}{18 r^2} \bigg[
      16r - 217 r^2 + 781 r^3 + 175 r^4 - 273 r^5 - 2 r^6
      \nonumber
      \\
      &&
     + 2 (8 + 28 r + 221 r^2 + 293 r^3 + 419 r^4 - 29 r^5 - 76 r^6) \ln(1 - r)
      \nonumber
      \\
      &&
     + 4r^3 (21 - 108 r + 61 r^2 + 38 r^3) \ln(r)
      \nonumber
      \\
      &&
     - 32r^2 (11 + 43 r + 9 r^2 - 4 r^3 + 16 r^4) \ln \left(\frac{\mu}{m_Q}\right)
      \nonumber
      \\
      &&
     + 6r^2 (7 - 35 r + 3 r^2 + 15 r^3 + 10 r^4) \Big(\pi^2 - 2 \Li_2(1 - r) + 2 \Li_2(r) \Big)
      \bigg]
      \bigg[\frac{1}{1-r}\bigg]_{+}
      \nonumber
      \\
      &&
      - \frac{1}{9} \bigg[
       159 + 88\pi^2 - 520\ln \left(\frac{\mu}{m_Q}\right) + 768\ln^2 \left(\frac{\mu}{m_Q}\right)
      \bigg]
      \delta(-1 + r)\,,
      \nonumber
      \\
      &&
 \\
 \mathcal{C}_{\rho_D}^{\rm NLO,\,A} &=&
      \frac{1}{27r^2}\bigg[
       12 r - 67 r^2 - 182 r^3 + 117 r^4 - 418 r^5 - 296 r^6
             \nonumber
      \\
      &&
      + 6 (2 + 7r + 28r^2 - 21r^3 + 110r^4 - 38r^5 - 28r^6) \ln(1 - r)
            \nonumber
      \\
      &&
      - 6 r^3 (3 - 18 r + 28 r^2 - 28 r^3) \ln(r)
            \nonumber
      \\
      &&
      + 36 r^2 (2 - 10 r + 9 r^2 - 13 r^3 + 18 r^4) \ln\left(\frac{\mu}{m_Q}\right)
            \nonumber
      \\
      &&
      - 9 r^2 (1 + r + 6r^2 - 8 r^3) \Big(\pi^2 - 2 \Li_2(1 - r) + 2 \Li_2(r) \Big)
      \bigg]
      \bigg[\frac{1}{1-r}\bigg]_{+}
      \nonumber
      \\
      &&
      - \frac{1}{108}\bigg[
      5381 - 360\pi^2
      - 5256\ln \left(\frac{\mu}{m_Q}\right)
      + 1584 \ln^2\left(\frac{\mu}{m_Q}\right)
      \bigg]
      \delta(-1 + r)\,.
      \nonumber
      \\
      &&
\end{eqnarray}
Again, the coefficient of the total rate is obtained from Eq.~(\ref{Citot}) by integrating the coefficient of the differential rate
over $r$ in the whole range. In the massive lepton case we obtain
\begin{eqnarray}
 C_{\rho_D}^{\rm LO} &=&
  \frac{1}{3} (\eta -1)(5 \eta ^3-43 \eta ^2+29\eta + 45)
 + 4(1-4 \eta) \eta^2 \ln (\eta )
 \nonumber
 \\
 &&
   + 32 (\eta +1) (\eta-1)^2 \bigg( \ln (1-\eta) - \ln \left(\frac{\mu }{m_Q}\right) \bigg)\,,
   \label{CD:totLO}
 \\
 C_{\rho_D}^{\rm NLO,\,F} &=&
 \frac{1}{216} \left(6 \eta ^4-14546 \eta^3+22967 \eta ^2+5338 \eta -13765\right)
 \nonumber
 \\
 &&
   -\frac{1}{54} \pi ^2 \left(30 \eta ^4-356 \eta ^3-192 \eta ^2+144 \eta
   +147\right)
    \nonumber
 \\
 &&
 + \frac{1}{108} \eta  \left(57 \eta ^3-6868 \eta ^2+4710 \eta +996\right) \ln (\eta )
    \nonumber
 \\
 &&
   -\frac{1}{108} \left(57 \eta ^4-9976 \eta ^3+15768 \eta ^2+888 \eta -6737\right) \ln (1-\eta )
       \nonumber
 \\
 &&
   +\frac{8}{9} \left(8 \eta ^4-155 \eta ^3+216 \eta ^2+31 \eta -100\right) \ln
   \left(\frac{\mu }{m_Q}\right)
    \nonumber
 \\
 &&
   +\frac{16}{3} (14-19 \eta ) \eta ^2 \ln
   (\eta ) \ln \left(\frac{\mu }{m_Q}\right)
    \nonumber
 \\
 &&
   -\frac{16}{3} (\eta +1) (\eta-1)^2 \left(9 \ln ^2(1-\eta )-25 \ln (1-\eta ) \ln
   \left(\frac{\mu }{m_Q}\right)+16 \ln ^2\left(\frac{\mu }{m_Q}\right)\right)
       \nonumber
 \\
 &&
   - \frac{1}{9}\left( 15\eta^4 - 4\eta^3 - 60\eta^2 + 132\eta - 83\right) \ln (1-\eta) \ln (\eta )
    \nonumber
 \\
 &&
   - \frac{1}{9} \left( 30\eta^4 + 772\eta^3 - 480\eta^2 - 48\eta - 47\right) \Li_2(\eta )
    \nonumber
 \\
 &&
   -\frac{20}{9} \eta ^2 (2 \eta -3) \left(\pi ^2 \ln (\eta )
   +9\Li_3(\eta )-3 \ln (\eta ) \Li_2(\eta )-9 \zeta (3)\right)\,,
 \\
 C_{\rho_D}^{\rm NLO,\,A} &=&
 - \frac{1}{108} \left( 296\eta^4 + 55\eta^3 - 1055\eta^2 + 1189\eta - 485\right)
 -\frac{1}{27} \pi ^2 \left(41\eta^3 - 69\eta - 24\right)
     \nonumber
 \\
 &&
 +\frac{1}{27} \left( 42 \eta^3 + 812\eta ^2-441 \eta +6\right) \eta  \ln (\eta )
     \nonumber
 \\
 &&
   -\frac{2}{27} \left(21 \eta ^4+571 \eta ^3-525 \eta ^2-591 \eta +524\right)\ln (1-\eta)
    \nonumber
 \\
 &&
   +\frac{2}{9} \left(27 \eta ^4+57 \eta^3-139 \eta ^2-61 \eta +116\right) \ln \left(\frac{\mu }{m_Q}\right)
   + \frac{40}{3} \eta ^3 \ln (\eta ) \ln \left(\frac{\mu }{m_Q}\right)
    \nonumber
 \\
 &&
   +\frac{4}{3} (\eta-1)^2 (\eta +1) \left(5 \ln
   ^2(1-\eta )+6 \ln (1-\eta ) \ln \left(\frac{\mu }{m_Q}\right)-11 \ln ^2\left(\frac{\mu
   }{m_Q}\right)\right)
       \nonumber
 \\
 &&
   +\frac{2}{9} \left(11 \eta ^3-33 \eta ^2+21 \eta +1\right) \ln (1-\eta ) \ln (\eta )
    \nonumber
 \\
 &&
   +\frac{2}{9} \left(115 \eta ^3-162 \eta ^2-15 \eta +10\right) \Li_2(\eta )
    \nonumber
 \\
 &&
   +\frac{8}{9} \eta ^3
   \left(\pi ^2 \ln (\eta )+9 \Li_3(\eta )-3 \ln (\eta ) \Li_2(\eta )-9 \zeta (3)\right)\,,
\end{eqnarray}
where $\Li_3(x)$ is the trilogarithm and $\zeta(x)$ is the Riemann zeta function.
In the massless lepton ($\eta=0$) case we obtain
\begin{eqnarray}
 C_{\rho_D}^{\rm LO} &=& -15 - 32 \ln \left(\frac{\mu }{m_Q}\right)\,,
 \label{CD:totLO0}
 \\
 C_{\rho_D}^{\rm NLO,\,F} &=&
  -\frac{13765}{216}
  -\frac{49 \pi^2}{18}
  -\frac{800}{9} \ln \left(\frac{\mu }{m_Q}\right)
  -\frac{256}{3} \ln ^2\left(\frac{\mu }{m_Q}\right)\,,
 \\
 C_{\rho_D}^{\rm NLO,\,A} &=&
  \frac{485}{108}
 +\frac{8 \pi^2}{9}
 +\frac{232}{9} \ln \left(\frac{\mu }{m_Q}\right)
 -\frac{44}{3} \ln ^2\left(\frac{\mu }{m_Q}\right)\,.
\end{eqnarray}
The LO coefficients given in Eqs. (\ref{CD:totLO}) and (\ref{CD:totLO0})
agree with~\cite{Lenz:2020oce,Mannel:2020fts,Moreno:2020rmk}. The NLO results are new.

\section{Evanescent operators}
\label{sec:ev}

The coefficient functions presented in Sec.~[\ref{sec:4qO}] are expressed in terms of the operator basis
\begin{eqnarray}
 \mathcal{O}_3^{hl} &=& (\bar h_v \gamma_{\mu} P_L T^a q)(\bar q \gamma^\mu P_L T^a h_v)\,,
 \\
 \mathcal{O}_4^{hl} &=& (\bar h_v P_L T^a q)(\bar q P_R T^a h_v)\,,
 \\
 \mathcal{O}_5^{hl} &=& (\bar h_v \gamma_\mu \gamma_\nu P_L T^a q)(\bar q \gamma^\mu \gamma^\nu P_R T^a h_v)\,,
 \\
 \mathcal{O}_6^{hl} &=& (\bar h_v \gamma_\mu \gamma_\nu \gamma_\alpha P_L T^a q)(\bar q \gamma^\mu \gamma^\nu \gamma^\alpha P_L T^a h_v)\,.
\end{eqnarray}
In $D=4$ the operators $\mathcal{O}_{5,6}^{hl}$ are redundant and they can be straightforwardly reduced to the
operators $\mathcal{O}_{3,4}^{hl}$ hereby reducing the number of operators in the basis. However, this is no longer true for arbitrary
$D$. Despite of this, it is possible to make a closer connection to a four-dimensional basis by choosing an operator basis with evanescent operators~\cite{Buras:1989xd,Dugan:1990df,Herrlich:1994kh,Beneke:2002rj,Mannel:2020fts}
\begin{eqnarray}
  \mathcal{O}_{3E}^{hl} &=& (\bar h_v \gamma_{\mu} P_L T^a q)(\bar q \gamma^\mu P_L T^a h_v)\,,
 \\
 \mathcal{O}_{4E}^{hl} &=& (\bar h_v P_L T^a q)(\bar q P_R T^a h_v)\,,
 \\
 E_1^{hl} &=& (\bar h_v \gamma_\mu \gamma_\nu \gamma_\alpha P_L T^a q) (\bar q \gamma^\mu \gamma^\nu \gamma^\alpha P_L T^a h_v)
 - (16-a\epsilon)(\bar h_v \gamma_\mu P_L T^a q) (\bar q \gamma^\mu P_L T^a h_v)\,,
 \\
 E_2^{hl} &=& (\bar h_v \gamma_\mu \gamma_\nu P_L T^a q) (\bar q P_R \gamma^\mu \gamma^\nu T^a h_v)
 - (4-b\epsilon)(\bar h_v P_L T^a q)(\bar q P_R T^a h_v)\,,
\end{eqnarray}
where $E_{1,2}^{hl}$ are the so-called evanescent operators and $a$, $b$ are arbitrary numbers which makes the choice of the evanescent operators ambiguous. This ambiguity is connected to the freedom in the choice of the renormalization scheme.
Eventually, the scheme dependence of the Wilson coefficients must cancel against the scheme dependence of the
matrix elements of the corresponding operators. It is conventional to use $a=4$ and $b=-4$ with $D=4-2\epsilon$.
This choice is motivated by the preservation of Fierz symmetry at one-loop order~\cite{Buras:1989xd,Beneke:2002rj}, and we will
refer to this choice as the {\it canonical} basis of four-quark operators.
In this basis only matrix elements of $\mathcal{O}_{3E,4E}^{hl}$ are non-zero, whereas matrix elements
of evanescent operators vanish, showing in this way a close connection to $D=4$.
The two basis are related by
\begin{eqnarray}
  \mathcal{O}_{3}^{hl} &=& \mathcal{O}_{3E}^{hl}\,,
 \\
 \mathcal{O}_{4}^{hl} &=& \mathcal{O}_{4E}^{hl}\,,
 \\
 \mathcal{O}_5^{hl} &=& E_2^{hl} + (4-b\epsilon)\mathcal{O}_{4,E}^{hl} \,,
 \\
 \mathcal{O}_6^{hl} &=& E_1^{hl} + (16-a\epsilon)\mathcal{O}_{3,E}^{hl} \,.
\end{eqnarray}
In the new basis the imaginary part of the transition operator in differential form becomes
\begin{equation}
 \mbox{Im}\, \mathcal{T} = \Gamma^0 |V_{qQ}|^2 \int_\eta^1 dr
 \bigg(\ldots
 + \mathcal{C}_{3E}^{hl} \frac{\mathcal{O}_{3E}^{hl}}{4m_Q^3}
 + \mathcal{C}_{4E}^{hl} \frac{\mathcal{O}_{4E}^{hl}}{4m_Q^3}
 + \mathcal{C}_{E_1}^{hl} \frac{E_1^{hl}}{4m_Q^3}
 + \mathcal{C}_{E_2}^{hl} \frac{E_2^{hl}}{4m_Q^3}
 \bigg)\,,
\end{equation}
where ellipses stand for the terms in Eq.~(\ref{eq:HQE-1-dq}) excluding
the operators $\mathcal{O}_i^{hl}$ ($i=3,\ldots,6$). The corresponding contribution to the decay width changes to
\begin{eqnarray}
\label{hqewidth2ev}
\Gamma(H_Q\rightarrow X_q \ell \bar \nu_\ell )
  &=& \Gamma^0 |V_{qQ}|^2 \int_\eta^1 dr
 \bigg[ \ldots
 	+ \mathcal{C}_{3E}^{hl} \frac{\rho_{3E}^{hl\,3}}{2m_Q^3}
	+ \mathcal{C}_{4E}^{hl} \frac{\rho_{4E}^{hl\,3}}{2m_Q^3}
	+ C_{E_1}^{hl} \frac{\rho_{E_1}^{hl\,3}}{2m_Q^3}
	+ C_{E_2}^{hl} \frac{\rho_{E_2}^{hl\,3}}{2m_Q^3}
 \bigg]\,,
\end{eqnarray}
where ellipses stand for the terms in Eq.~(\ref{hqewidth2}) excluding
the matrix elements $\rho_i^{hl\,3}$ ($i=3,\ldots,6$). The new matrix elements are defined by following Eq.~(\ref{ME4q}).
The relation between the coefficients of the differential rate in the two basis reads
\begin{eqnarray}
 \mathcal{C}_{3E}^{hl} &=& \mathcal{C}_{3}^{hl} + (16-a\epsilon)\mathcal{C}_{6}^{hl}\,,
 \\
 \mathcal{C}_{4E}^{hl} &=& \mathcal{C}_{4}^{hl} + (4-b\epsilon)\mathcal{C}_{5}^{hl}\,,
 \\
 \mathcal{C}_{E_2}^{hl} &=& \mathcal{C}_{5}^{hl}\,,
 \\
 \mathcal{C}_{E_1}^{hl} &=& \mathcal{C}_{6}^{hl}\,.
\end{eqnarray}
Note that the only coefficients which depend on the numbers $a$, $b$ parametrizing the ambiguity on the definition of the evanescent operators are $\mathcal{C}_{3E}^{hl}$ and $\mathcal{C}_{4E}^{hl}$.

The operator mixing between four-quark operators changes to
\begin{eqnarray}
 \delta \mathcal{C}_1^{hl\,\rm \overline{MS}} &=& 3\mathcal{C}_1^{hl\,B} C_F \frac{\alpha_s}{4\pi}\frac{1}{\epsilon}\,,\quad\quad
 \delta \mathcal{C}_2^{hl\,\rm \overline{MS}} = 3\mathcal{C}_2^{hl\,B} C_F \frac{\alpha_s}{4\pi}\frac{1}{\epsilon}\,,\quad\quad
 \delta \mathcal{C}_{3E}^{hl\,\rm \overline{MS}} = -3\mathcal{C}_1^{hl\,B} \frac{\alpha_s}{4\pi}\frac{1}{\epsilon}\,,
  \nonumber
 \\
  \delta \mathcal{C}_{4E}^{hl\,\rm \overline{MS}} &=& -3\mathcal{C}_2^{hl\,B} \frac{\alpha_s}{4\pi}\frac{1}{\epsilon}\,,\quad\quad
 \delta \mathcal{C}_{E_1}^{hl\,\rm \overline{MS}} = - \frac{1}{4}\mathcal{C}_1^{hl\,B} \frac{\alpha_s}{4\pi}\frac{1}{\epsilon}\,,\quad\quad
 \delta \mathcal{C}_{E_2}^{hl\,\rm \overline{MS}} = - \frac{1}{4}\mathcal{C}_2^{hl\,B} \frac{\alpha_s}{4\pi}\frac{1}{\epsilon}\,,
\end{eqnarray}
which is independent of $a$, $b$. For the differential rate, a different choice of $a$, $b$ corresponds to the following shift in the coefficients of the four-quark operators
\begin{eqnarray}
  \mathcal{C}_{3E}^{hl}(a_1,b_1) - \mathcal{C}_{3E}^{hl}(a_2,b_2)
 &=& \frac{\alpha_s}{\pi}16\pi^2 (1 - \eta)^2 (2 + \eta) (a_1 - a_2) \delta(-1 + r)\,,
 \\
 \mathcal{C}_{4E}^{hl}(a_1,b_1) - \mathcal{C}_{4E}^{hl}(a_2,b_2)
 &=&
  - \frac{\alpha_s}{\pi} 32 \pi^2 (1 - \eta)^2 (1 + 2 \eta) (b_1 - b_2) \delta(-1 + r)\,.
\end{eqnarray}
For the {\it canonical} choice of evanescent operators we obtain
\begin{eqnarray}
 \mathcal{C}_{3E}^{hl}(4,-4) &=&
 -\frac{\alpha_s}{\pi}64\pi^2 \bigg\{
  \frac{2}{3r^3} (r - \eta)^2
   ( 2\eta + r(1 - 11\eta)- r^2 (25 + 2 \eta)
     \nonumber
  \\
  &&
  + 2 r^3 (1 + \eta) + 4r^4)
  \bigg[\frac{1}{1-r}\bigg]_{+}
  \nonumber
  \\
  &&
  + (1 - \eta)^2 \bigg[
   13 + 8 \eta - 6(2 + \eta)\bigg(\ln(1 - \eta) - \ln\left(\frac{\mu}{m_Q}\right) \bigg)
  \bigg]
  \delta(-1 + r)
  \bigg\}\,,
  \\
  \mathcal{C}_{4E}^{hl}(4,-4) &=&
 \frac{\alpha_s}{\pi}128 \pi^2 \bigg\{
 -\frac{2}{3r^3} (r - \eta)^2 (8\eta + r (4 + 7 \eta) + r^2 (2 + 7 \eta)
   \nonumber
  \\
  &&
 + r^3 (11 - 4 \eta) - 8r^4)
  \bigg[\frac{1}{1-r}\bigg]_{+}
  \nonumber
  \\
  &&
  + (1 - \eta)^2 \bigg[
   8 + 13 \eta - 6 (1 + 2 \eta)\bigg( \ln(1 - \eta) - \ln\left(\frac{\mu}{m_Q}\right) \bigg)
  \bigg] \delta(-1 + r)
  \bigg\}\,.
\end{eqnarray}
For the total rate, a different choice of $a$, $b$ corresponds to the following shift in the coefficients of the four-quark operators
\begin{eqnarray}
 C_{3E}^{hl}(a_1,b_1) - C_{3E}^{hl}(a_2,b_2)
 &=& \frac{\alpha_s}{\pi} 16 \pi^2  (\eta -1)^2 (\eta +2) (a_1-a_2)\,,
 \\
 C_{4E}^{hl}(a_1,b_1) - C_{4E}^{hl}(a_2,b_2)
 &=& - \frac{\alpha_s}{\pi} 32 \pi^2 (\eta -1)^2 (2 \eta +1) (b_1-b_2)\,.
\end{eqnarray}
For the {\it canonical} choice of evanescent operators we obtain
\begin{eqnarray}
 C_{3E}^{hl}(4,-4) &=&
 -\frac{\alpha_s}{\pi} \frac{64}{9} \pi^2 \bigg[
  (\eta -1)(110 \eta^2 - 7\eta - 205)
 + 6 (11 \eta +6) \eta ^2 \ln (\eta)
 \nonumber
 \\
 &&
 - 54 (\eta - 1)^2 (\eta + 2)\bigg( \ln (1-\eta ) - \ln \left(\frac{\mu}{m_Q}\right)\bigg)
   \bigg]\,,
   \\
C_{4E}^{hl}(4,-4) &=& \frac{\alpha_s}{\pi}\frac{128}{9} \pi^2
 \bigg[
  211 \eta^3 -315\eta^2 + 9\eta +95
 + 12 \eta ^2 (11 \eta -12) \ln (\eta )
 \nonumber
 \\
 &&
 -54 (\eta -1)^2 (2 \eta +1) \bigg( \ln (1-\eta ) - \ln \left(\frac{\mu}{m_Q}\right)\bigg)
   \bigg]\,.
\end{eqnarray}
The two expressions above are in agreement with~\cite{Beneke:2002rj,Lenz:2013aua}. The massless lepton
case can be straightforwardly obtained from the equations above by taking $\eta=0$.

Let us discuss now the effect that the choice of an operator basis with evanescent operators has over the Darwin coefficient. 
The one-loop operator mixing of four-quark operators with the Darwin operator in Eq.~(\ref{CDmix3}) changes to
\begin{eqnarray}
 \delta \mathcal{C}_D^{\rm \overline{MS},\,4q\,(1)} &=&
 - \bigg( 2\mathcal{C}_{1,\,B}^{hl} - \mathcal{C}_{2,\,B}^{hl} + ( 2\mathcal{C}_{3E,\,B}^{hl} - \mathcal{C}_{4E,\,B}^{hl} + 2a\epsilon \mathcal{C}_{E_1,\,B}^{hl} - b\epsilon \mathcal{C}_{E_2,\,B}^{hl}
  )\left(C_F - \frac{C_A}{2}\right)\bigg)\frac{1}{96\pi^2\epsilon}\bar{\mu}^{-2\epsilon}\,.
 \nonumber
 \\
 &&
 \label{ad1l3qodev}
\end{eqnarray}
Note that the bare coefficients $C_{E_1,\,B}^{hl}$ and $C_{E_2,\,B}^{hl}$ contain a $1/\epsilon$ pole
which cancels the explicit $\epsilon$ in front of them in Eq.~(\ref{ad1l3qodev}). The anomalous dimension gives an extra $1/\epsilon$ factor generating in this way a contribution 
due to operator mixing of evanescent operators with the Darwin operator.
However, the whole contribution $\delta \mathcal{C}_D^{\rm \overline{MS},\,4q\,(1)}$ remains unaltered. In particular,
the explicit dependence on $a$, $b$ in Eq.~(\ref{ad1l3qodev}) cancels against the dependence on $a$, $b$ of the
coefficients $\mathcal{C}_{3E,\,B}^{hl}$ and $\mathcal{C}_{4E,\,B}^{hl}$. In other words, we find that the Darwin coefficient is
independent of the choice of the evanescent operators.

\section{Numerical estimates}
\label{sec:disc}
In this section we evaluate the numerical impact of the new results by inserting illustrative values for the non-perturbative matrix elements and the parameters entering in the Wilson coefficients. The numerical values are provided in tables~\ref{tab:me} and \ref{tab:par}. Note that there is a tension between the numerical value of $\rho_D$ for $B$-decays estimated from \cite{Bordone:2021oof} and \cite{Bernlochner:2022ucr} with a 
factor 6-9 difference. We take the largest value in order to estimate an upper bound
for the size of these corrections.
\begin{table}
 \begin{center}
\begin{tabular}{|c|c|c|c|c|c|c|}
\hline
  & $B^{+}$ & $B^0$ & $B_s$ & $D^{+}$ & $D^0$ & $D_s$ \\
   \hline
$\rho_1^{hl\,3}$ (GeV$^3$)& $0.0107$ & $2.77\cdot 10^{-5}$ & $3.98\cdot 10^{-5}$ & $1.37\cdot 10^{-5}$ & $1.37\cdot 10^{-5}$ &
$8.73\cdot 10^{-3}$\\
 $\rho_2^{hl\,3}$ (GeV$^3$)& $0.0106$ & $-1.92\cdot 10^{-5}$ & $-2.86\cdot 10^{-5}$ & $-1.01\cdot 10^{-5}$ & $-1.01\cdot 10^{-5}$ & $8.70\cdot 10^{-3}$ \\
 $\rho_{3E}^{hl\,3}$ (GeV$^3$)& $-1.76\cdot 10^{-4}$ & $-4.26\cdot 10^{-6}$ &  $-6.36\cdot10^{-6}$ & $-2.38\cdot 10^{-6}$ & $-2.38\cdot 10^{-6}$ & $- 9.06\cdot 10^{-5}$ \\
 $\rho_{4E}^{hl\,3}$ (GeV$^3$) & $-4.26\cdot 10^{-6}$ & $3.19\cdot 10^{-6}$ & $4.77\cdot 10^{-6}$ & $1.79\cdot 10^{-6}$ &
 $1.78\cdot 10^{-6}$ & $-8.71 \cdot 10^{-7}$\\
 $\rho_D^3$ (GeV$^3$) & 0.408 & 0.408 & 0.607 & 0.174 & 0.174 & 0.256 \\
 \hline
\end{tabular}
\caption{Numerical values for the matrix elements used to estimate the size of corrections. 
The values are taken from~\cite{King:2021xqp,Lenz:2022rbq} and translated to our notation by using the numerical values for the meson masses and decay constants also provided 
in these references.}
\label{tab:me}
\end{center}
\end{table}
\begin{table}
 \begin{center}
\begin{tabular}{|c|c|c|c|}
\hline
 Parameter & Numerical value & Parameter & Numerical value\\
   \hline
 $\mu$ & $m_Q$ & $m_\tau$ & $1.777$ GeV \\
 $m_b$ & $4.7$ GeV &  $\alpha_s(m_b)$ & $0.217$ \\
 $m_c$& $1.6$  GeV& $\alpha_s(m_c)$ &  $0.340$ \\
 \hline
\end{tabular}
\caption{Numerical values of parameters entering in the Wilson coefficients.}
\label{tab:par}
\end{center}
\end{table}
We use the HQE in the {\it canonical} basis
of four-quark operators and consider the total rate and moments of the
$q^2$-distribution, which are defined in analogy to the HQE of the width by
\begin{eqnarray}
 \label{hqemoments}
M_n(H_Q\rightarrow X_q \ell \bar{\nu}_\ell ) &=&
\int_{\eta}^{1}dr\, r^n \frac{d\Gamma(H_Q \rightarrow X_q \ell \bar{\nu}_\ell )}{dr}
 \nonumber
\\
&=&
\Gamma^0 |V_{qQ}|^2
 \bigg[   M_{n,0}
- M_{n,\mu_\pi}\frac{\mu_\pi^2}{2m_Q^2}
+ M_{n,\mu_G}\frac{\mu_G^2}{2m_Q^2}
- M_{n,\rho_D} \frac{\rho_D^3}{2m_Q^3}
 - M_{n,\rho_{LS}} \frac{\rho_{LS}^3}{2m_Q^3}
 \nonumber
 \\
 &&
 + M_{n,1}^{hl} \frac{\rho_1^{hl\,3}}{2m_Q^3}
 + M_{n,2}^{hl} \frac{\rho_2^{hl\,3}}{2m_Q^3}
 + M_{n,3E}^{hl} \frac{\rho_{3E}^{hl\,3}}{2m_Q^3}
 + M_{n,4E}^{hl} \frac{\rho_{4E}^{hl\,3}}{2m_Q^3}
 \bigg]
  \, .
\end{eqnarray}
Note that in the expression above we have already neglected matrix elements of evanescent operators and
that the zeroth moment corresponds to the total width. The
coefficients of the moments are related to the coefficients of the spectrum by
\begin{eqnarray}
 M_{n,i}(\eta) &=& \int_{\eta}^{1}dr\, r^n \mathcal{C}_i(r,\eta)\,.
 \label{Mitot}
\end{eqnarray}
The low $q^2$ is difficult to detect and experimentalists use cuts and integrate up to the available
$q^2$~\cite{Belle:2021idw}. Moments with low cuts can be obtained from the expression above by integrating in the
desired range. For simplicity we consider the width and the first two moments without cuts. In tables~\ref{tab:m0}, \ref{tab:m1} and \ref{tab:m2} 
we compare the size of the Darwin and four-quark terms at LO and NLO relative to the leading power term.

\begin{table}
 \begin{center}
\begin{tabular}{|c|c|c|c|c|c|c|c|c|c|}
\hline
 $n=0$ & $B^{+}\tau \bar{\nu}_\tau$ & $B^0\tau \bar{\nu}_\tau$ & $B_s\tau \bar{\nu}_\tau$ & $B^{+}e \bar{\nu}_e$ & $B^0 e \bar{\nu}_e$ & $B_s e \bar{\nu}_e$ & $D^{+}e \bar{\nu}_e$ & $D^0 e \bar{\nu}_e$ & $D_s e \bar{\nu}_e$ \\
   \hline
$\rho_D^3$ (LO)& $12\%$ & $12\%$ & $17\%$  & $4\%$ & $4\%$ & $5\%$ & $43\%$ & $43\%$ & $63\%$\\
$\rho_D^3$ (NLO)& $3 \%$ & $3\%$ & $5\%$ & $1\%$ & $1\%$ & $2\%$ & $25\%$ & $25\%$ & $37\%$ \\
 $\rho_i^{hl\,3}$ (LO)& $13\%$ & $-0.3\%$ & $-0.5\%$ & $-0.1\%$  &  $-0.1\%$ & $-0.2\%$ & $-2\%$ & $-2\%$ &  $-3\%$\\
  $\rho_i^{hl\,3}$ (NLO)& $-0.9\%$ & $0.07\%$ & $0.1\%$ & $0.1\%$  & $0.03\%$ & $0.04\%$ & $0.7\%$ & $0.7\%$  & $3\%$\\
 \hline
\end{tabular}
\caption{Relative contribution of the Darwin operator and four-quark operators to the leading term for the total width. }
\label{tab:m0}
\end{center}
\end{table}
We observe that corrections due to the Darwin term are in general more important for semitauonic decays than for semileptonic decays
of $B$ mesons. For the $B$ semitauonic decay width the LO and NLO represent $\sim 15\%$ and $\sim5\%$ corrections, respectively.
For the $B$ semileptonic decay width the LO and NLO represent $\sim 5\%$ and $\sim 1\%$ corrections, respectively.
For semileptonic $D$ decays the corrections due to the Darwin term are very large, endangering the convergence of the HQE.
For example, at LO and NLO they represent $\sim 50\%$ and $\sim 30\%$ correction to the width.

As for moments of the spectrum, the convergence of the HQE worsens for higher moments. The reason is that moments enhance
the region of the phase space where perturbation theory and the HQE break down. The convergence is better
for semileptonic decays than for semitauonic decays of $B$ mesons and for lower moments. Considering that the value of
$\rho_D$ must be clarified and that it might become a factor 6-9 smaller, the convergence of the first moments in
$B$ decays seems to be good, but it is unclear to what extent without clarifying the value of $\rho_D$.
In general terms, the $1/m_b^3$ corrections can be safely used to improve the precision of the HQE for
the $B\rightarrow X_u \ell \bar{\nu}_\ell$ decay. On the contrary, the Darwin term of moments for $D$ decays becomes even larger
than the the leading term pointing out the breakdown of the HQE, and therefore that the $1/m_c^3$ corrections can not be used to improve the precision 
in the $D$-meson spectrum.

Finally, we observe that in general the NLO corrections to the Darwin term are rather large, as they correspond
to $\sim20\%$ correction to the coefficient.

We also observe that the semitauonic decay of $B^{+}$ receives large corrections from four-quark operators unlike it happens
in the other decays. The reason is that, due to the tau mass, the four-quark operators
$\mathcal{O}_{1,2}^{hl}$ do not combine in perpendicular
form $\mathcal{O}_{\perp}^{hl}$, unlike it happens in the massless case.
The matrix elements of such perpendicular combinations are very much suppressed. Also matrix elements
of octet operators or of operators involving different spectator quarks in the operator and the state are suppressed. In particular, all these matrix elements are exactly zero in VIA. Therefore,
the HQE of the $B^{+}$ semitauonic decay is the only one where $\mathcal{O}_{1,2}^{hl}$ do not combine in $\mathcal{O}_{\perp}^{hl}$ and whose matrix elements involve the same spectator quark in the operators and the state.
For the decay width the LO and NLO terms represent
$\sim 13\%$ and $\sim -0.9\%$ corrections, respectively. For moments the corrections become larger as it happens with the
Darwin term. In general, deviations from VIA give rise to very small corrections compared to the Darwin term. Therefore
we predict the $B^{+}\rightarrow X_u \tau \bar{\nu}_\tau$ decay width to be $\sim 10\%$ larger than the
$B^{0}\rightarrow X_u \tau \bar{\nu}_\tau$ decay. This observation still has to be confirmed by experiment due to its measurement is very challenging.

Overall, we find the Darwin term to be the dominant dimension 6 contribution except in the semitauonic decay of $B^{+}$, where
four-quark operators give a similar contribution.
\begin{table}
 \begin{center}
\begin{tabular}{|c|c|c|c|c|c|c|c|c|c|}
\hline
 $n=1$ & $B^{+}\tau \bar{\nu}_\tau$ & $B^0\tau \bar{\nu}_\tau$ & $B_s\tau \bar{\nu}_\tau$ & $B^{+}e \bar{\nu}_e$ & $B^0 e \bar{\nu}_e$ & $B_s e \bar{\nu}_e$ & $D^{+}e \bar{\nu}_e$ & $D^0 e \bar{\nu}_e$ & $D_s e \bar{\nu}_e$ \\
   \hline
$\rho_D^3$ (LO)& $38\%$ & $38\%$ & $57\%$  & $23\%$  & $23\%$ & $34\%$ & $280\%$  & $280\%$ & $410\%$  \\
$\rho_D^3$ (NLO)& $9\%$ & $9\%$ & $14\%$ & $6\%$  & $6\%$ & $8\%$ & $106\%$ & $106\%$ & $155\%$ \\
 $\rho_i^{hl\,3}$ (LO)& $28\%$ & $-0.7\%$ & $-1\%$ & $-0.4\%$ & $-0.4\%$ & $-0.7\%$ & $-6\%$ & $-6\%$ & $-9\%$  \\
  $\rho_i^{hl\,3}$ (NLO)& $-2\%$ & $0.2\%$ & $0.2\%$ & $0.5\%$  & $0.1\%$ & $0.1\%$ & $2.3\%$ & $2.3\%$ & $12\%$ \\
 \hline
\end{tabular}
\caption{Relative contribution of the Darwin operator and four-quark operators to the leading term for the first moment. }
\label{tab:m1}
\end{center}
\end{table}
\begin{table}
 \begin{center}
\begin{tabular}{|c|c|c|c|c|c|c|c|c|c|}
\hline
 $n=2$ & $B^{+}\tau \bar{\nu}_\tau$ & $B^0\tau \bar{\nu}_\tau$ & $B_s\tau \bar{\nu}_\tau$ & $B^{+}e \bar{\nu}_e$ & $B^0 e \bar{\nu}_e$ & $B_s e \bar{\nu}_e$ & $D^{+}e \bar{\nu}_e$ & $D^0 e \bar{\nu}_e$ & $D_s e \bar{\nu}_e$ \\
   \hline
$\rho_D^3$ (LO)& $90\%$ & $90\%$ & $135\%$   & $67\%$ & $67\%$  & $100\%$ & $808\%$ & $808\%$  & $1185\%$ \\
$\rho_D^3$ (NLO)& $21\%$ & $21\%$ & $31\%$ & $15\%$ & $15\%$ & $23\%$ & $288\%$  & $288\%$  & $423\%$ \\
 $\rho_i^{hl\,3}$ (LO)& $53\%$ & $-1\%$ & $-2\%$ & $-1\%$  & $-1\%$ & $-1\%$ & $-14\%$  & $-14\%$ & $-20\%$ \\
  $\rho_i^{hl\,3}$ (NLO)& $-3\%$ & $0.3\%$ & $0.4\%$ & $1\%$ & $0.2\%$ & $0.3\%$ & $5\%$ & $5\%$  & $30\%$\\
 \hline
\end{tabular}
\caption{Relative contribution of the Darwin operator and four-quark operators to the leading term for the second moment.}
\label{tab:m2}
\end{center}
\end{table}


\section*{Conclusions}

In this work we have presented analytical results for the $\alpha_s$ corrections to the coefficients of the Darwin operator and four-quark operators appearing at order $1/m_Q^3$ in the HQE of the $H_Q\rightarrow X_q \ell \bar{\nu}_\ell$ decay for both, the total width and the spectrum on the dilepton invariant mass in the case of a massless quark and both a massless ($\ell= e,\,\mu$) or massive ($\ell= \tau$) lepton in the final state. The results can be applied to the 
CKM suppressed $B\rightarrow X_u \ell\bar{\nu}_\ell$ decay or, to some extent, to the CKM favoured $D\rightarrow X \ell\bar{\nu}_\ell$ ($\ell\neq \tau$) decay.

We have observed that the newly computed NLO corrections to the Darwin term are rather large, as they typically correspond
to $\sim 20\%$ correction to the Darwin term at LO. For the semitauonic and semileptonic decay rate of $B$ mesons the NLO corrections represent $\sim5\%$ and $\sim 1\%$ corrections
to the leading term. For the semileptonic decay rate of $D$ mesons they correspond to $\sim 25\%$ correction to the leading term, showing a
much slower convergence of the HQE for $D$ than for $B$. 

The convergence of the HQE worsens for higher moments.
For $B$ mesons it is crucial to clarify the value of $\rho_D^3$ in order to make a clear statement about the convergence of the HQE for
higher moments. For moments of the $D$ meson spectrum the convergence is very bad, pointing out that the $1/m_c^3$ corrections can not be used to improve the precision of the HQE.

We conclude that the $1/m_Q^3$ corrections can be used to improve the precision of the
HQE for the $B\rightarrow X_u \ell \bar{\nu}_\ell$ decay rate and the first few moments and, to less extent,
for the $D\rightarrow X \ell \bar{\nu}_\ell$ decay rate.

We have also observed that, unlike what it happens in the other decay channels, the semitauonic decay width and moments of $B^{+}$ receive large corrections from four-quark operators and that they are similar in size that the Darwin term.
In particular, we expect the $\Gamma(B^{+}\rightarrow X_u \tau \bar{\nu}_\tau) \sim 0.1\Gamma(B^{0}\rightarrow X_u \tau \bar{\nu}_\tau)$. This prediction still has to be confirmed by the experiment.

Overall, we find the Darwin term to be the dominant dimension 6 contribution except in the semitauonic decay of $B^{+}$, where
four-quark operators give a similar contribution.

The main application of our results is for the background subtraction of the $B\rightarrow X_u \ell \bar\nu_\ell$ decay in the
measurement of $B\rightarrow X_c \ell \bar\nu_\ell$ decay, used for the precise extraction of $|V_{cb}|$ from $q^2$-moments
and the precise measurement of $R(D^{(*)})$. Other important
applications are for the lifetimes of $B$ and $D$ hadrons, the study of the $D$-hadron spectrum 
as well as the extraction of the ratio $|V_{ub}/V_{cb}|$.
A rigorous phenomenological analysis updating the predictions for the different observables is left to future publications.

Finally, the computation carried out in this paper represents a step towards the computation of the Darwin coefficient
at NLO for the non-leptonic decay width, which is sought at present~\cite{Lenz:2022rbq,Mannel:2023zei}.
In particular, the current work can be used to understand how the operator mixing works when there is also two-loop mixing with
four-quark operators in a much simpler scenario than for non-leptonic decays, where a large proliferation of four-quark operators occurs.

\subsection*{Acknowledgments}
I thank M.~Fael, G.~Finauri, P.~Gambino and K.~Vos for pointing out inconsistencies in the relation between the chromomagnetic operator and the spin-orbit operator coefficients in~\cite{Moreno:2022goo,Mannel:2021zzr}.
I also thank S.~Kollatzsch for her interest in this work and for comments on the manuscript.
The early stage of this project has received funding from the Deutsche Forschungsgemeinschaft
(DFG, German Research Foundation) under grant 396021762 - TRR 257 
``Particle Physics Phenomenology after the Higgs Discovery''. The final stage of this project has 
received funding from the European Union’s Horizon 2020
research and innovation program under the Marie Skłodowska-Curie grant
agreement No.~884104 (PSI-FELLOW-III-3i).

\end{document}